%% file: main.tex
\pdfoutput=1

\documentclass[10pt,pageno,letterpaper,twocolumn,10pt]{article}

\usepackage{usenix}

\input{pkgs}
\input{cmds}

\usepackage{hyperref}

\titlespacing*{\section}{0pt}{0.3\baselineskip}{0.3\baselineskip}
\titlespacing*{\subsection}{0pt}{0.2\baselineskip}{0.2\baselineskip}

\begin{document}

\setlength{\pdfpageheight}{\paperheight}
\setlength{\pdfpagewidth}{\paperwidth}

\input{hdr}

\date{}
\maketitle

\input{abstract}

\input{intro}

\input{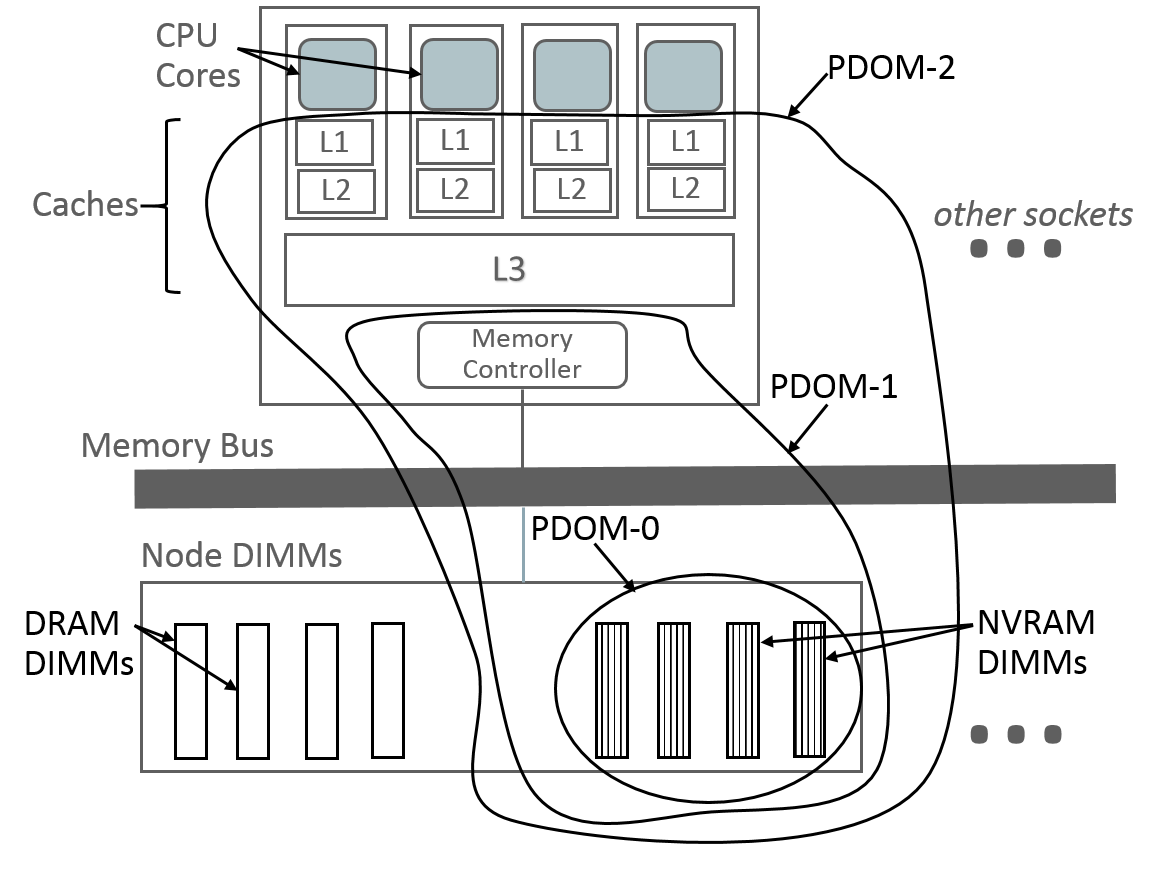}
\input{impl}

\input{mem-mgmt}

\input{perf}

\input{related-work}

\input{conclusion}

\bibliographystyle{abbrv}
\bibliography{refs}

\end{document}

%% file: pkgs.tex
\usepackage[normalem]{ulem}
\usepackage{fullpage}
\usepackage{amsmath}
\usepackage{graphicx}
\usepackage{booktabs}
\usepackage{verbatim}
\usepackage{multirow}
\usepackage{fancyvrb}
\usepackage{microtype}
\usepackage[T1]{fontenc}
\usepackage{balance}
\usepackage[labelfont=bf,font=small,skip=1pt]{caption}
\usepackage{pythontex}
\usepackage{titlesec}

%% file: cmds.tex
\input{code/fmt}

\newcommand{\load}{\textsf{load}}
\newcommand{\store}{\textsf{store}}
\newcommand{\nop}{\textsf{nop}}

\newcommand{\pmalloc}{\textsf{pm\_alloc}}
\newcommand{\pmfree}{\textsf{pm\_free}}

\newcommand{\memcached}{\textsf{mem\-cached}}

\newcommand{\invalid}{\textsf{IDLE}}
\newcommand{\running}{\textsf{RUN\-NING}}
\newcommand{\committed}{\textsf{COM\-MIT\-TED}}
\newcommand{\aborted}{\textsf{ABORT\-ED}}

\newcommand{\txnbegin}{\textsf{TXN\_BEGIN}}

\newcommand{\txnwrite}{\textsf{TXN\_WRITE}}
\newcommand{\txnread}{\textsf{TXN\_READ}}
\newcommand{\txnopen}{\textsf{TXN\_OPEN}}

\newcommand{\memcpy}{\textsf{mem\-cpy}}

\newcommand{\clwb}{\textsf{clwb}}
\newcommand{\clflushopt}{\textsf{clflush-opt}}
\newcommand{\clflush}{\textsf{clflush}}
\newcommand{\pcommit}{\textsf{pcommit}}
\newcommand{\sfence}{\textsf{sfence}}
\newcommand{\cas}{\textsf{compare-and-swap}}

\newcommand{\kvget}{\textsf{get}}
\newcommand{\kvput}{\textsf{put}}

\newcommand{\ra}[1]{\renewcommand{\arraystretch}{#1}}

%% file: hdr.tex
\vspace{-1.0in}
\title{Persistent Memory Transactions\thanks{Work done when all authors were affiliated with Oracle.  Contact: \textsf{virendra.marathe@oracle.com}}}
\vspace{-1.5in}

\author{Virendra J. Marathe$^1$ \hspace{0.05in} Achin Mishra$^2$ \hspace{0.05in} Amee
  Trivedi$^3$ \hspace{0.05in} Yihe Huang$^4$ \hspace{0.2in} Faisal Zaghloul$^5$ \\
  Sanidhya Kashyap$^6$ \hspace{0.05in} Margo Seltzer$^{1,4}$ \hspace{0.05in}
  Tim Harris \hspace{0.05in} Steve Byan$^1$ \hspace{0.05in} Bill Bridge$^7$
  \hspace{0.05in} Dave Dice$^1$ \\\\
  $^1$Oracle Labs \hspace{0.05in} $^2$IBM \hspace{0.05in}
  $^3$University of Massachusetts, Amherst \hspace{0.05in}
  $^4$Harvard University \\ $^5$Yale University \hspace{0.05in}
  \hspace{0.05in} $^6$Georgia Institute of Technology \hspace{0.05in}
  $^7$Oracle Corporation
  }

\vspace{-1.5in}

%% file: abstract.tex
\begin{abstract}
  
This paper presents a comprehensive analysis of performance trade offs
between implementation choices for transaction runtime systems on
persistent memory.  We compare three implementations of transaction
runtimes: \emph{undo logging}, \emph{redo logging}, and
\emph{copy-on-write}.  We also present a memory allocator that plugs
into these runtimes.  Our microbenchmark based evaluation focuses on
understanding the interplay between various factors that contribute to
performance differences between the three runtimes -- read/write
access patterns of workloads, size of the \emph{persistence domain}
(portion of the memory hierarchy where the data is effectively
persistent), cache locality, and transaction runtime bookkeeping
overheads.  No single runtime emerges as a clear winner.  We confirm
our analysis in more realistic settings of three ``real world''
applications we developed with our transactional API: (i) a key-value
store we implemented from scratch, (ii) a SQLite port, and (iii) a
persistified version of \memcached, a popular key-value store.  These
findings are not only consistent with our microbenchmark analysis, but
also provide additional interesting insights into other factors
(e.g. effects of multithreading and synchronization) that affect
application performance.

\end{abstract}

%% file: intro.tex
\vspace{0.1in}
\section{Introduction}
\label{sec:intro}

Byte-addressable persistent memory technologies (e.g.
\emph{spin-transfer torque MRAM (STT-MRAM)}~\cite{hosomi05,huai08},
\emph{memristors}~\cite{strukov08}), that approach the performance of
DRAM (100-1000x faster than state-of-the-art NAND flash) are coming,
as evidenced by Intel and Micron Technologies' recent announcement of
their 3D XPoint persistent memory technology~\cite{3dxpoint}.
While the simple \load{}/\store{} based interface of these technologies
is appealing, it introduces new challenges; a simple \texttt{store}
does not immediately persist data, because processor state and various
layers or the memory hierarchy (e.g., store buffers, caches) are
expected to remain \emph{nonpersistent} for the foreseeable future.
Prior research~\cite{condit09,joshi15,pelley14} and processor vendors,
such as Intel, have proposed new hardware
instructions~\cite{intel-isa} to flush or write cache lines back to
lower layers in the memory hierarchy and new forms of \emph{persist
  barrier} instructions that can be used to order persistence of
stores.  However, even with these new instructions, correctly writing
programs to use them remains a daunting task.
\autoref{fig:pointer-example} illustrates this challenge -- the
programmer must carefully reason about the order in which updates to
various pieces of the application's persistent data structures are
persisted.  Omission of even a single flush, write back, or persist
barrier instruction can result in persistent data inconsistencies in
the face of failures.

\begin{figure}
  \centering
  \small
  \input{code/pointer.c}
  \vspace{-0.1in}
\caption{Example illustrating complexities of programming with just
  the hardware instructions for persisting data. \textsf{p} is a
  pointer hosted in persistent memory.
  \textsf{clone} clones its argument object (\textsf{obj}).  The
  programmer must persist the clone before \textsf{p}'s assignment,
  otherwise an untimely failure could result in a state where the
  clone is not persisted but \textsf{p}'s new value is persisted.}
\vspace{-0.25in}
\label{fig:pointer-example}
\end{figure}

These programming challenges have resulted in investigations of
transactional mechanisms to access and manipulate data on persistent
memory~\cite{bridge15,chakrabarti14,coburn11,giles15,kolli16,pmem-io,volos11}.
However, we observe that research on transaction runtime
implementations in this new context is in its early stages.  In
particular, researchers and practitioners appear to have favored or
rejected an implementation strategy based on intuition (e.g. redo
logging is a bad idea since redo log lookups for read-after-write
scenarios are
expensive~\cite{bridge15,chakrabarti14,coburn11,kolli16,pmem-io}) or
inadequate evaluation (e.g. evaluation only on write-heavy
workloads~\cite{giles15}).  No one to our knowledge has endeavored to
do a comprehensive analysis of performance trade offs between these
implementations over a wide swath of workload, enclosing system, and
transaction runtime choice parameters.  This paper presents a holistic
approach toward understanding these performance trade offs.

We consider the implications of \emph{persistence
domains}~\cite{snia13} on persist barrier overheads
(see~\autoref{sec:pdomain}).  Briefly, a persistence domain is the
portion of the memory hierarchy that is considered to be ``effectively
persistent'' -- the underlying hardware/software system ensures that
data that reaches its persistence domain is written to the persistent
media before the system is shut down, either planned or due to
failures.  We introduce a new taxonomy of persistence domain choices
enabled by different hardware systems.

We present our three transaction runtimes based on \emph{undo
logging}, \emph{redo logging} and \emph{copy-on-write} implementations
of transactional writes (see~\autoref{sec:txns}).  We also present our
memory management algorithm that plugs into all three runtimes
(see~\autoref{sec:mem-mgmt}).  All our runtimes, including the memory
manager, have been optimized to reduce the number of persist barriers
required to commit a transaction.

Our microbenchmarking (see~\autoref{sec:perf}), performed on Intel's
Software Emulation Platform~\cite{dulloor14,zhang15}, comprehensively
sweeps through read-write mix ratios within transactions and shows how
performance trends in the transaction runtimes change as the
read-write mix within transactions changes, and over a wide range of
persist barrier latencies.  Our analysis reveals the significant
influence of a combination of factors -- read/write mix, transaction
runtime specific bookkeeping overheads, persist barrier latencies, and
cache locality -- which determines the performance of a runtime.
Because of the interplay of these factors, no single runtime's
performance dominates the rest in all settings.  We find similar
performance trade offs in three ``real world'' workloads: (i) a
key-value store we developed from scratch, (ii) a port of SQLite, and
(iii) a port of \memcached.  The benchmarks provide insights in
additional factors that influence performance (e.g. effects of
multithreading and synchronization, overheads in other parts of the
application).

%% file: code/pointer.c.tex
\begin{Verbatim}[commandchars=\\\{\},codes={\catcode`\$=3\catcode`\^=7\catcode`\_=8}]
\PY{k+kt}{void} \PY{o}{*}\PY{n}{p}\PY{p}{;} \PY{c+c1}{// pointer to persistent memory}
\PY{p}{.}\PY{p}{.}\PY{p}{.}
\PY{c+c1}{// obj and clone(obj) are persistent}
\PY{n}{p} \PY{o}{=} \PY{n}{clone}\PY{p}{(}\PY{n}{obj}\PY{p}{)}\PY{p}{;}
\end{Verbatim}

%% file: pdomain.tex
\section{Persistence Domain}
\label{sec:pdomain}

\begin{figure}[t]
\includegraphics[width=0.8\columnwidth,height=4.5cm]{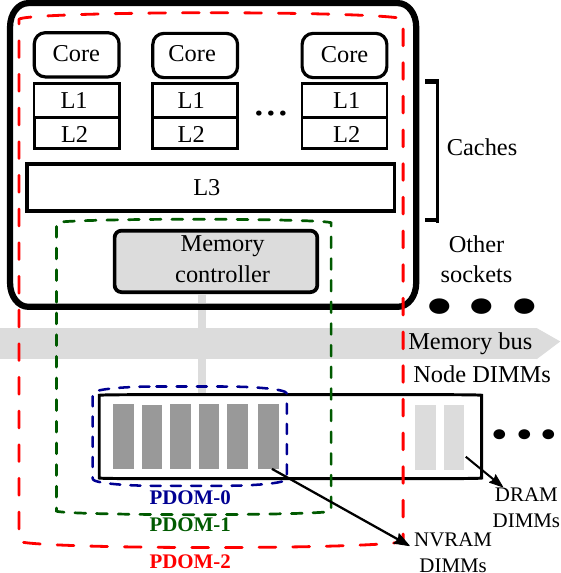}
\centering
\caption{Persistence domains of a near future processor socket that
  hosts persistent memory DIMMs.}
\label{fig:pdomain}
\end{figure}

While data hosted in persistent memory DIMMs is expected to survive
power failures, the rest of the memory hierarchy (e.g. processor
caches, memory controller buffers, etc.) is fundamentally not
persistent.  However, system solutions do exist that make various
parts of the memory hierarchy ``effectively persistent''.  For
instance, in battery backed
systems~\cite{chen96,izraelevitz16,narayanan12,nawab15}, where the
processor state and caches can be flushed out to persistent memory
DIMMs on power failure, the whole memory hierarchy effectively becomes
persistent.  Another example is the \emph{asynchronous DRAM refresh
  (ADR)} feature provided by modern processors, where the memory
controller buffers are flushed out to memory DIMMs on power
failure~\cite{intel-nvdimm-driver-guide}.  With the ADR feature, the
memory controller buffers can be considered effectively persistent
since the data is guaranteed, discounting ADR hardware failures, to
persist.  There may be other ways to slice the memory hierarchy in
persistent and nonpersistent parts; however, we focus on 3 specific
partitioning strategies that we believe will capture most future
system configurations.

A \emph{persistence domain}~\cite{snia13} as the portion of memory
hierarchy where data is effectively persistent.  As shown in
\autoref{fig:pdomain}, we classify persistence domains in three
categories: (i) PDOM-0, which contains only the persistent memory
DIMMs.  (ii) PDOM-1, which includes PDOM-0 and memory controller
buffers. Modern processors with ADR capabilities and persistent memory
DIMMs effectively support PDOM-1.  (iii) PDOM-2, which includes the
entire memory hierarchy as well as processor state, such as store
buffers, containing persistent data.  Battery backed systems support
PDOM-2.

\begin{table}
  \ra{0.8}
  \centering
  \small
  \input{figures/tbl-pdoms}
\caption{Persistent memory primitives needed on Intel's upcoming
  processors~\cite{intel-isa} for different persistence domains.}
\label{table:primitives}
\end{table}

The persistence domain affects the instruction sequence needed to
persist updates.
\autoref{table:primitives} depicts the instructions needed to
persist these updates on (near future) Intel processors with
persistent memory~\cite{intel-isa}. There are two phases to the
persistent update process:
(i) The actual write (i.e., \store) and (ii) the persist barrier.
PDOM-0 and PDOM-1 require a flush instruction in addition to the
\store\ to move data into the persistence domain.
Both the \clwb\ and \clflushopt\ trigger asynchronous cache-line sized
writes to the memory controller; they differ in that \clflushopt\
invalidates the cache line while \clwb\ does not.
In principle, the flush instructions can be delayed, and almost
certainly should be for multiple \store\ instructions to the same
cache line. In practice, as they are asynchronous, starting the writeback
sooner speeds up the persist barriers in the second phase of this process.
In PDOM-2, flush instructions are not needed, since store buffers
and caches are part of the persistence domain.

In PDOM-0, the persist barrier needs to ensure that all flushes have
completed (the first \sfence), and then force any updates in the memory
controller to be written to the DIMMs (\pcommit). As the \pcommit\ is
asynchronous, persistence requires the second \sfence\ to indicate when the
\pcommit\ has completed.
In PDOM-1, the persist barrier need only ensure that prior flushes
have completed, since the memory controller now resides inside the
persistence domain.
PDOM-2 requires no further action as data is persisted as soon as it has
been \store{}d.
Intel has recently
deprecated the \pcommit\ instruction~\cite{intel-nvdimm-driver-guide}.
However, we include it in our discussion as a
concrete example of a PDOM-0 persistence domain.
Note that \clwb, \clflushopt, and
\pcommit\ have \store\ semantics in terms of memory ordering, and
applications must take care to avoid problematic reordering of loads
with these instructions, using \sfence\ or other instructions with
fence semantics.

%% file: figures/tbl-pdoms.tex
{\footnotesize
\begin{tabular}{llll}
\toprule
\multirow{2}{*}{\textbf{Operations}}
& \multicolumn{3}{c}{\textbf{Persistence domains}}
\\\cmidrule(lr){2-4}
& \multicolumn{1}{c}{\textsc{PDOM-0}}
& \multicolumn{1}{c}{\textsc{PDOM-1}}
& \multicolumn{1}{c}{\textsc{PDOM-2}} \\
\midrule
\multirow{2}{*}{\textsf{Writes}}
& \textsf{store}
& \textsf{store}
& \textsf{store} \\
& \textsf{clwb/clflush-opt}
& \textsf{clwb/clflush-opt}
& \\ \hline
\multirow{3}{*}{\textsf{Ordering}}
& \textsf{sfence}
& \textsf{sfence}
& \textsf{nop} \\
\multirow{3}{*}{\textsf{persists}}
& \textsf{pcommit}
&
& \\
& \textsf{sfence}
&
& \\
\bottomrule
\\
\end{tabular}

}

%% file: impl.tex
\section{Persistent Transactions}
\label{sec:txns}

Similar to prior works~\cite{bridge15,chakrabarti14,
  coburn11,volos11}, our programming model is based on the
abstractions of persistent regions, persistent data types, and
transactions.  A persistent region is a contiguous portion of the
application's address space populated by a memory mapped file that is
hosted in persistent memory.  The region can be accessed with the
\load{}/\store\ interface, hosts a heap, and a user instantiated root
pointer.  The heap provides \pmalloc\ and \pmfree\ functions, callable
only from transactions.

\begin{figure}
  \centering
  \small
  \input{code/counter.c}
\caption{Example of a simple transaction that increments a counter in
  a persistent object.}
\label{fig:interface}
\end{figure}

Our work focuses on supporting the needs of skilled system software
programmers, who require programming support for just \emph{failure
  atomicity} -- across failure boundaries, either all updates of
transactions persist or none of them do.  These programmers are adept
at manually using synchronization techniques to avoid data races in
concurrent settings.  Hence, while various semantic models for
persistent memory transactions have been explored -- full ACID
transactions in the spirit of transactional memory
~\cite{coburn11,volos11}, failure-atomic critical
sections~\cite{chakrabarti14}, and failure-atomic
transactions~\cite{bridge15,giles15,kolli16,pmem-io} -- we focus on
failure atomic transactions.  Our programming model, implemented as a
C library, supports transaction begin and commit operations, as well
as transactional accessor macros.  We also provide macros to define
persistent types, which act as wrappers around traditional data types.
\autoref{fig:interface} provides a brief example illustrating our
basic API.  We also provide common memory buffer operators such as
\textsf{memcpy}, \textsf{memcmp}, and \textsf{memset}

Our transactional accessors introduce runtime overheads.  For
programmers that want to elide these overheads without compromising
correctness of their applications, we provide the \textsf{PM\_UNWRAP}
macro that returns a pointer to the data type instance wrapped by a
persistent type instance.  This tool can be useful in special
circumstances including accessing objects in read-only mode, and
initializing newly allocated objects.

The primary focus of our work is on understanding the performance trade
offs between different implementations of persistent memory transactions.
To that end we have developed
three different transaction runtime systems: (i) undo
logging, (ii) redo logging, and (iii) copy-on-write (COW).
All the runtimes store transaction metadata in a
persistent data structure called the \emph{transaction descriptor},
which is assigned to a thread as part of \txnbegin.
A descriptor is always in one of
four states: \invalid, \running, \aborted, or
\committed.  A descriptor that is not in use is in the
\invalid\ state. \txnbegin\ transitions the descriptor
into the \running\ state.
A transaction commits by entering the
\committed\ state and aborts by entering the
\aborted\ state.
After the runtime cleans up a descriptor's internal
state and buffers, the descriptor returns to the
\invalid\ state.  During its execution, a transaction may read, write,
allocate, and deallocate persistent objects using our API.

\begin{figure}[t]
\includegraphics[width=\columnwidth]{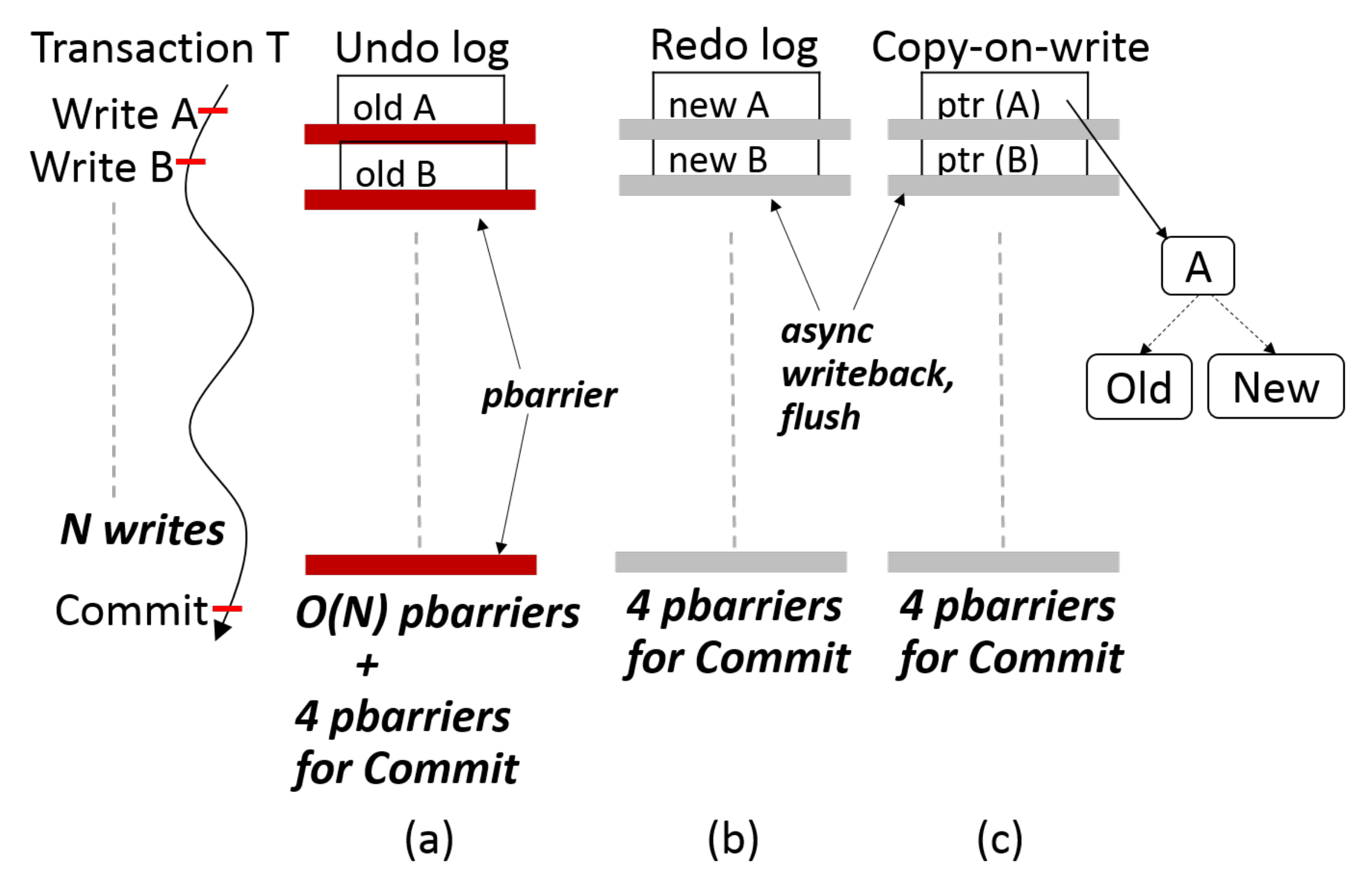}
\centering
\caption{Transaction write implementations.  Transaction $T$ applies
  $N$ distinct writes.  \textsf{pbarrier} is a potentially high
  latency persist barrier, \textsf{async wb/flush} are asynchronous
  cache line write-back or flush instructions respectively.}
\label{fig:txns}
\end{figure}

\subsection{Undo Log based Transactions}

\autoref{fig:txns}\textsf{(a)} shows transaction $T$'s undo logging
activities.  The undo log is implemented as a simple chunked list.
Transaction $T$ writes $w$, using \txnwrite\, producing a
log record containing the original value of $w$.
This log record must be persisted before $w$ is modified.
A typical implementation of the undo log append
requires two persist barriers -- one to persist the new log
record and one to change the log's tail pointer.
Although correct, this approach leads to 2$N$ persist
barriers for $N$ appends, which results in high overheads.  

Our implementation requires only a single persist barrier per record.
Instead of relying on a correct tail pointer during recovery, we infer
the tail of the log.
We assign each transaction a monotonically increasing persistent 64-bit
version number.
Each undo record contains a transaction version
number, a 64-bit checksum, and a 64-bit \emph{prolog} sentinel value
that appears at the beginning of the undo record.
So, we write the prolog sentinel value, the transaction's version
number, the record contents, and then compute and write the
checksum.
Then we issue a single persist barrier.
If a failure occurs before or during execution of
the persist barrier, and only part of the undo record has been persisted,
we will detect a checksum mismatch during recovery.
We also maintain the log tail pointer, but update it
\emph{after} the persist barrier, so the tail update is
guaranteed to persist on or before the next record write and persist.
Recovery can rely on the following invariant: the tail
pointer will be at most one record behind the actual end of log.
So log recovery requires only that we examine the record after the
current end of log and determine if there is a valid log record present.

One of the most compelling benefits of undo logging is that transactional
reads are implemented as uninstrumented
loads~\cite{coburn11,bridge15,kolli16,pmem-io,volos11}.

An undo log transaction commits in four steps: (i) First it ensures that
all transactional writes are persisted, which requires
one persist barrier. (ii) Then it logically commits the transaction by
appending the commit record to the transaction's undo log.  (It also
switches the transaction's state to \committed, but that does not have
to persist.)  Steps (iii) and (iv) are largely related to
transactional metadata cleanup, which requires persistence only if 
the transaction allocated or deallocated persistent memory
(see \autoref{sec:mem-mgmt}).
(iii) Persist the allocation/deallocation calls' effects
and cleans up the transaction's metadata.
(iv) Mark the transaction \invalid;
this state change needs to be persisted only if the
transaction did allocations/deallocations.

\subsection{Redo Log based Transactions}
\label{sec:redo-logging}

\autoref{fig:txns}\textsf{(b)} shows transaction $T$'s redo logging
activities.  Like the undo log, the redo log is implemented as a
simple chunked list.
Transaction $T$ writes $w$, using \txnwrite\, producing a log record
containing the new value of $w$.  The record need not persist at the
time of the write; if a failure occurs, the entire redo log can be
discarded.  However, an implementation, like ours, may proactively
schedule a low latency asynchronous writeback/flush of the record.

The challenge for redo logging schemes is handling read-after-write
accesses.  As the new value appears only in the log, a subsequent read
must consult the log for the latest value.  A naive implementation
could walk down the whole log looking for the latest value.
Furthermore, the redo log lookup could be done for \emph{every}
subsequent read done by the transaction.  The resulting overhead can
be significant.  We apply two optimizations to overcome these
overheads.

First, we add a 64-bit bitmap to each persistent object's header to
indicate current writers of that object.  A writer first sets its
corresponding bit during the write (the bit is cleared after the
transaction completes).  A read checks the object's writers bitmap to
determine if a redo log lookup is necessary, and does on if so.  If a
lookup is not necessary, the read becomes an uninstrumented \load.
Each transaction maps to a unique bit in the object's writers bitmap.
Up to 64 transactions can concurrently ``own'' a bit in the writers
bitmap.  This can be easily extended to larger bitmaps, but at present
we force additional transactions to consult the redo log for all their
reads.

Second, we avoid scanning the entire log by maintaining a
per-transaction hash table indexed by persistent object base
addresses.  The hash table record points to the latest redo log record
for that object.  Each such record also contains a pointer to the
previous redo log record for that object, if one exists.  We
effectively superimpose a linked stack of records for each object
within the redo log.  This avoids unnecessary traversal of unrelated
log records during a redo log lookup.

Committing a transaction requires persisting the redo log.
After the persist completes, the
transaction logically commits by updating its state to
\committed, and then persists the new state with a second
persist barrier.  After the logical commit, the runtime
applies the redo log to each modified object 
and issues a third persist barrier.
Finally, we mark the transaction \invalid, and
persist it. In total, the redo logging implementation requires
four persist barriers for commit, but none on abort.

\subsection{Copy-on-Write based Transactions}

Our copy-on-write (COW) implementation introduces an extra level of
indirection between a persistent type instance (the wrapper) and the
real data type instance (payload) it encloses.  As shown in
\autoref{fig:txns}\textsf{(c)}, the persistent type contains pointers
to \emph{old} and \emph{new} versions of the enclosed type's
instances.
Before modifying an object, a transaction creates a new copy of the
payload.
We provide a special
\txnopen\ API that applications can use to obtain read-only or
read-write access to a persistent object:

\begin{verbatim}
  TXN_OPEN(txn, obj, mode, copy_ctor);
\end{verbatim}

\noindent where \textsf{mode} is either read-only or read-write, and
\textsf{copy\_ctor} is the copy constructor.  The copy constructor can
be used to clone specialized objects (e.g. linked structures,
self-relative pointers, etc.).  A NULL copy constructor will default
to using \memcpy.

Each transaction descriptor maintains a \emph{write set} containing
the list of objects the transaction has written.
Objects are added to the write set in \txnopen\ invocations
with read-write mode.  Object wrappers also contain the writing transaction's
ID (assuming at most one writer per persistent object), which is used to direct
transactional reads to the appropriate payload copy.

Payload copies, as well as writes to their wrappers, need not
be persisted during the writer's transaction.
The transaction's write
set and the objects it writes to are persisted using a single
persist barrier at the beginning of the commit operation.
Then, the runtime updates the transaction's state to \committed\ and
persists it.

The post-commit cleanup requires four steps:
(i) make the modified (new) object payload the real (old) payload,
(ii) reset new to NULL,
(iii) discard (deallocate) the old payload, and
(iv) clear the writer's ID from the wrapper.
This process is susceptible to
memory leaks: a failure between steps (i) and (iii) can result in the
reference to the old payload being lost. We avoid this leak
by adding another field in the wrapper, called \textsf{old\_backup},
that is set to point to the old payload in \txnopen.
This update is
persisted during the first persist barrier in the commit operation.
\textsf{old\_backup} is used to deallocate the old payload.
Next, the transaction's
allocations/deallocations are all persisted. The third persist barrier
is issued after all this cleanup. Then, the transaction updates
its state to \invalid\ and persists it using a fourth persist
barrier.  This ensures that no further cleanup is needed.
Finally, we clear the transaction's ID from 
all the objects to which it wrote.
If a transaction aborts, only the last two clean up related persist
barriers are needed for correct rollback.

%% file: code/counter.c.tex
\begin{Verbatim}[commandchars=\\\{\},codes={\catcode`\$=3\catcode`\^=7\catcode`\_=8}]
\PY{k}{struct} \PY{n}{foo} \PY{p}{\PYZob{}}
    \PY{k+kt}{int} \PY{n}{cnt}\PY{p}{;} 
\PY{p}{\PYZcb{}}\PY{p}{;}
\PY{c+c1}{// pm\PYZus{}foo, the persistent version of type foo}
\PY{n}{DEFINE\PYZus{}PM\PYZus{}TYPE}\PY{p}{(}\PY{n}{foo}\PY{p}{)}\PY{p}{;} 
\PY{c+c1}{// x points to an instance of pm\PYZus{}foo    }
\PY{n}{pm\PYZus{}foo} \PY{o}{*}\PY{n}{x}\PY{p}{;} 
\PY{c+c1}{// failure\PYZhy{}atomic transaction for x\PYZhy{}\PYZgt{}cnt++;}
\PY{k+kt}{pm\PYZus{}txn\PYZus{}t} \PY{n}{txn}\PY{p}{;}
\PY{k}{do} \PY{p}{\PYZob{}}
    \PY{n}{TXN\PYZus{}BEGIN}\PY{p}{(}\PY{n}{txn}\PY{p}{)}\PY{p}{;}
    \PY{k+kt}{int} \PY{n}{counter}\PY{p}{;} \PY{c+c1}{// temporary}
    \PY{n}{TXN\PYZus{}READ}\PY{p}{(}\PY{n}{txn}\PY{p}{,} \PY{n}{x}\PY{p}{,} \PY{n}{cnt}\PY{p}{,} \PY{o}{\PYZam{}}\PY{n}{counter}\PY{p}{)}\PY{p}{;}  
    \PY{n}{counter}\PY{o}{+}\PY{o}{+}\PY{p}{;}
    \PY{n}{TXN\PYZus{}WRITE}\PY{p}{(}\PY{n}{txn}\PY{p}{,} \PY{n}{x}\PY{p}{,} \PY{n}{cnt}\PY{p}{,} \PY{o}{\PYZam{}}\PY{n}{counter}\PY{p}{)}\PY{p}{;} 
    \PY{n}{status} \PY{o}{=} \PY{n}{TXN\PYZus{}COMMIT}\PY{p}{(}\PY{n}{txn}\PY{p}{)}\PY{p}{;}         
\PY{p}{\PYZcb{}} \PY{k}{while} \PY{p}{(}\PY{n}{status} \PY{o}{!}\PY{o}{=} \PY{n}{TXN\PYZus{}COMMITTED}\PY{p}{)}\PY{p}{;}
\end{Verbatim}

%% file: mem-mgmt.tex
\section{Persistent Memory Management}
\label{sec:mem-mgmt}

Memory management is a foundational tier in any software stack.  We
anticipate that applications using transactions to access persistent
data will routinely allocate and deallocate persistent objects within
these transactions.  Most previous work on persistent memory
management focuses either on wear-leveling~\cite{moraru13} or
techniques for correct allocation that tolerates
failures~\cite{bridge15,coburn11,dulloor14}, disregarding the overhead
due to persist barriers.  Volos et al.~\cite{volos11} present an
algorithm that effectively eliminates persist barriers for memory
allocation/deallocation calls within a transaction.  But that works
only in their redo logging transactions.  Our algorithm is similar in
nature, but works with all of our transaction runtimes.

We build our algorithm on previous approaches that separate the
allocator's metadata in persistent and nonpersistent
halves~\cite{dulloor14,volos11}.  Our allocator is modeled after the
Hoard allocator~\cite{berger00}, where the heap is divided in shared
and thread-private superblocks.  Each superblock, hosted in persistent
memory, contains a persistent bitmap indicating allocation status of
corresponding blocks, and nonpersistent metadata (free and used lists)
hosted in DRAM.  A superblock is protected by a nonpersistent lock.

Each transaction maintains a persistent private
\emph{allocation log} that consists of all the allocation/deallocation
requests made by the transaction.  In a \pmalloc\ call, the
nonpersistent metadata of the superblock is updated by the transaction
and a corresponding record is appended to its allocation log.
\pmfree\ simply appends an entry to the allocation log.

The first persist barrier in a transaction's commit operation persists
the allocation log as well.  Once the transaction persists its
\committed\ state, operations in the allocation log are reflected in
the persistent metadata (bits are flipped using \cas\ instructions to
avoid data races, and then the cache lines are written back or
flushed).  The post-commit cleanup phase's first persist barrier
persists these flipped bits, and the last persist barrier marks the
transaction as \invalid.  Note that \pmfree\ calls' nonpersistent heap
metadata (free and used lists of a superblock) is updated \emph{after}
the cleanup persists.

%% file: perf.tex
\section{Empirical Evaluation}
\label{sec:perf}

Our performance evaluation comprises two parts: (i) microbenchmarking,
where we sweep through a comprehensive range of read/write mixes
within transactions to identify performance patterns of the
transaction runtimes over changing read/write proportions under varied
assumptions about persist barrier latencies; and (ii) evaluation of
three ``real-world'' applications -- a new persistent key-value store
we developed, a port of SQLite~\cite{sqlite} that uses our
transactions to persist the database, and a persistent version of
\memcached~\cite{memcached} -- that confirms, and adds to, our
findings reported in the micrbenchmarking part.

We conducted all our experiments on Intel's Software Emulation
Platform~\cite{dulloor14,zhang15}.  This emulator hosts a dual socket
16-core processor, with 512GB of DRAM.  384GB of that DRAM is
configured as ``persistent memory'' and 128GB acts as regular memory.
Persistent memory is accessible to applications via mmapping files
hosted in the PMFS instance~\cite{dulloor14} installed in the
emulator.

The emulator emulates the \clflushopt\ and \pcommit\ instructions.
The \clflushopt\ is implemented using the \clflush\ instruction.
Since \clflushopt\ evicts the target cache line, we expect it to lead
to significant increase in cache miss rates, thereby degrading
application performance; 2-10X in our microbenchmarking, which we do
not report in detailed due to space restrictions.  We therefore focus
our evaluation on the \clwb\ instruction, which does not evict the
target cache line, and is likely to be the instruction of choice for
applications on Intel platforms.  We emulate \clwb\ behavior with a
nop (it is not supported in the emulator); the actual latency of
persisting the data is incurred by the subsequent persist barrier,
which we emulate by using the emulator's \pcommit\ instruction.  We
note that Intel recently announced deprecation of the
\pcommit\ instruction~\cite{intel-nvdimm-driver-guide} from future
Intel processors; the persist barrier in that case is simply an
\sfence\ instruction (see~\autoref{table:primitives}).  However, in
the emulator, the \pcommit\ instruction simply stalls the calling
thread for a configurable amount of time~\cite{dulloor14,zhang15},
which lets us experiment with different latencies of the real persist
barrier -- it will be difficult to know the real latency of a persist
barrier until real hardware becomes available, so any evaluation needs
to target a broad range of latencies, which we do in our experiments.
Latency of \load{}s from persistent memory is a configurable parameter
as well.  \store\ latency in the emulator is the same as DRAM
\store\ latency.

We conducted experiments over a wide range of latency parameters for
persist barriers (0 -- 1000 nanoseconds) and \load{}s (100 -- 500
nanoseconds).  We report results for a \load\ latency of 300
nanoseconds, and 3 different persist barrier latencies -- 0, 100 and
500 nanoseconds, labeled as PDOM-2, PDOM-1, and PDOM-0 respectively to
map them to the different persistence domains from our taxonomy.
These latencies represent the overall performance trends we observed
over the broader range of latencies.

\subsection{Transaction Latency}

Our latency microbenchmarking focuses on understanding performance of
the transaction runtimes under different read/write loads.  The
synthetic Array microbenchmark developed by the SoftWrAP
work~\cite{giles15} suffices this purpose.  Array contains a
2-dimensional array of 64-bit integers hosted in persistent memory.
The first dimension contains 10 million slots (each is an array of
integers); the second dimension's (slot's) size is configurable, we
vary it from 1 (8 bytes) to 64 (512 bytes) entries.  Array
continuously runs transactions, each of which randomly accesses a
contiguous set of 20 slots.  The slot sizes cover a broad range of
access granularities.  Each slot access can be a simple read of all
the integers in the slot, or a read-write that increments all integers
in the slot.  We vary the number of slots accessed in read-only or
read-write mode for different test runs.  In addition, we implemented
two versions of slot writes: (i) a ``one shot'' update version, called
Array, where the transaction copies the slot integers in a private
(nonpersistent) buffer, increments the integers in that buffer and
then writes it back to the persistent slot; and (ii) a
``read-after-write intensive'' version, called Array-RAW, where each
integer in the slot is individually incremented ``in-place''.  The
second version helps us understand overheads related to
read-after-write accesses in the redo logging transaction runtime.  We
report results as the mean of three 10-second test runs preceded by a
10-second warmup phase (we observed less than 5\% deviation from the
mean in all results).  Array microbenchmark is single threaded and is
sufficient for latency measurements.

\begin{figure*}[t]
  \centering
  \input{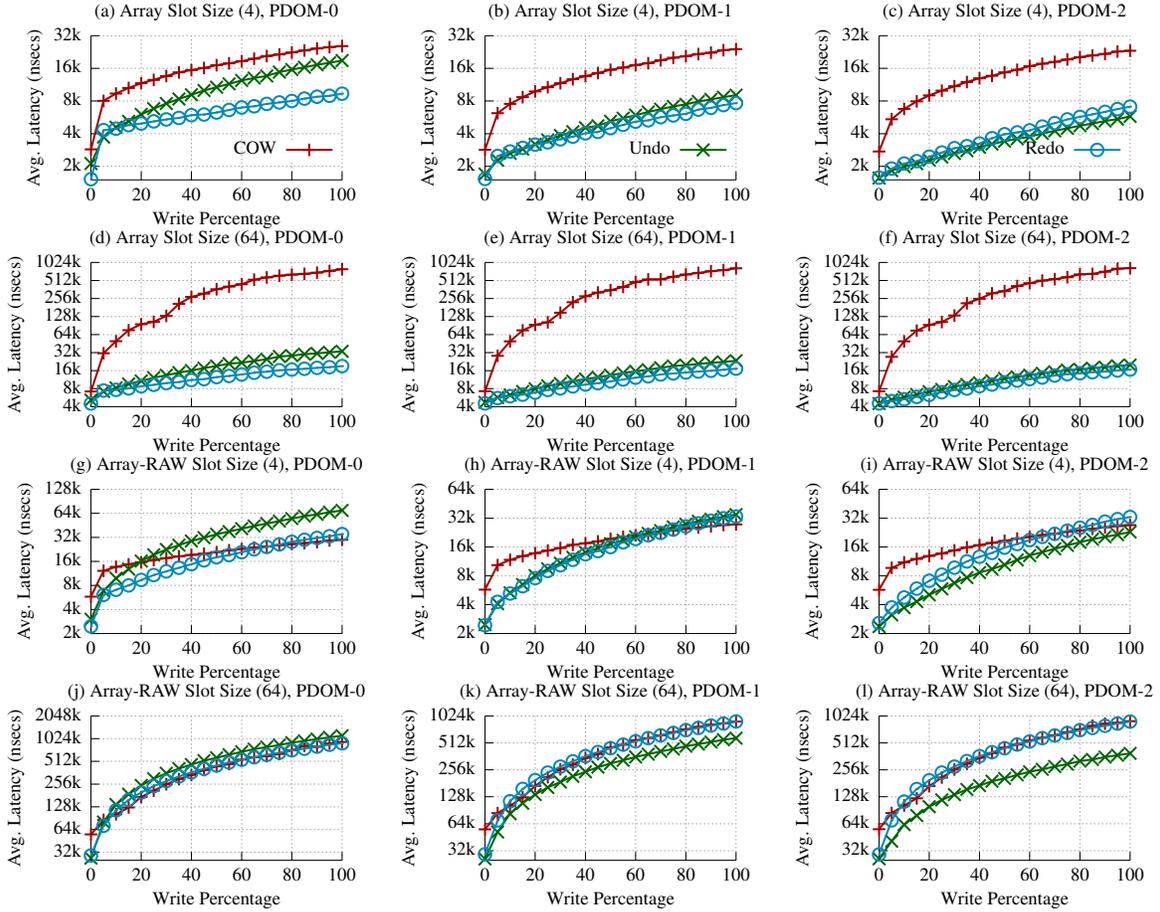}

  \vspace{0.25in}
  \caption{Average latency of transactions with increasing write
    percentage.  Y-axes are in log scale. Each transaction accesses 20
    slots.  X-axes start with read-only access runs; we progressively
    increase read-write accesses as we move right.}
  \label{fig:ubench-latency}
\end{figure*}

\autoref{fig:ubench-latency} shows latency results of our experiments
for slot sizes of 4 and 64 (they capture the performance patterns of
the configurations with other slot sizes we tested), and over the
three different persist barrier latencies discussed above.  Latency
graphs for Array appear in~\autoref{fig:ubench-latency}(a)--(f).  The
first takeaway of these is that COW performs worst across the board.
In fact the margin grows from 2X to 50X compared to the best
performing alternative when we move from 4-integer slots to 64-integer
slots.  The overheads of COW are largely related to worse cache
locality (proportionally high cache miss rates) compared to the undo
and redo logging alternatives -- (i) the extra level of indirection,
and (ii) the constant cloning of objects as they are updated.  This
gets worse with increasing granularity of objects.

As expected, undo logging performance degrades with increasing persist
barrier latencies.  This impact is most noticeable for PDOM-0, where
the persist barrier overheads are high -- the barrier overhead in undo
logging tends to increase the latency gap (up to 2X) with redo logging
as the percentage of writes per transaction increases
(\autoref{fig:ubench-latency}(a),(d)).  The same behavior manifests in
PDOM-1 configurations (\autoref{fig:ubench-latency}(b),(e)), albeit at
a lower scale (up to 20\% higher latency than redo logging), since the
persist barrier latencies are lower.

For PDOM-2 however, the persist barrier is a nop.  This shows
in\autoref{fig:ubench-latency}(c), where the slot size is 4.  Undo
logging either performs as well, or better than redo logging.  On
further observation, we realized that redo logging actually performs
increasingly worse than undo logging as the write percentage grows (up
to 25\% worse at 100\% writes).  Furthermore, at 0\% writes, redo and
undo logging are comparable in performance.  This implies that the
source of overheads is in the writes done by redo logging
transactions.  We determined that the operations related to
maintaining the lookup structure for accelerating read-after-write
lookups was the source of these overheads.
\autoref{fig:ubench-latency}(f) presents a contrasting result, where,
inspite of zero persist barrier latency, undo logging performs
increasingly worse (up to 15\%) than redo logging as the write
percentage per transaction grows.  We determined that the overheads in
undo logging were related to the checksum computations we needed to
avoid an extra persist barrier per undo log append.  Eliminating the
checksum brought undo and redo logging performance at parity in these
test runs.  This indicates an interesting trade off in performance of
undo logging implementations, where it is best to use checksums for
transactions that do fine grain writes, whereas it may be best to use
2 persist barriers per undo log append for transactions that make
coarse grained updates.

Notice that for read-only test runs (leftmost points in these graphs),
redo logging is either at parity with undo logging, or slightly better
(\autoref{fig:ubench-latency} (a)).  Our persistent object header
checks to detect read-after-write scenarios appear to have no effect
on redo logs performance.  In the absence of real read-after-write
scenarios, all the overhead in redo logging appears to be related to
writes that do the bookkeeping needed for fast read-after-write
lookups.

\autoref{fig:ubench-latency}(g)--(l) show performance of the
transaction runtimes on Array-RAW as the percentage of
read-after-write instances increases.  While the performance of COW
transactions remains more or less identical to their performance on
Array, both redo and undo logging transactions perform relatively
worse.  This is directly attributable to the proportional amount of
churn happening on the redo/undo logs (one log record per integer
increment).  For PDOM-0 test runs, the high persist barrier latency
combined with amplified number of persist barriers (4X or 64X) in undo
logging, leads to worst performance (even in comparison with COW
transactions) beyond a modest percentage of writes.  Redo logging, on
the other hand, incurs overheads related to read-after-write lookups
(through the transaction's lookup table and per-object update lists),
which are more modest than the persist barrier overheads, but are
nonethelees high enough to force redo logging perform as badly as COW
transactions for a modest percentage of writes.  At PDOM-1 persist
barrier latencies however, these lookups turn out to be relatively
more expensive, because of which undo logging performs comparably or
better than redo logging.  This difference increases furthermore for
PDOM-2 where the persist barrier is a nop.  These results affirm the
overall intuition of read-after-write lookup overheads in redo logging
in prior work~\cite{bridge15,chakrabarti14,coburn11,pmem-io}.  The
critical question of how often do such instances arise in real world
applications is something we address later in our evaluation.

\subsection{Memory allocation performance.}

\begin{figure}[t]
\centering
\vspace{0.05in}
\input{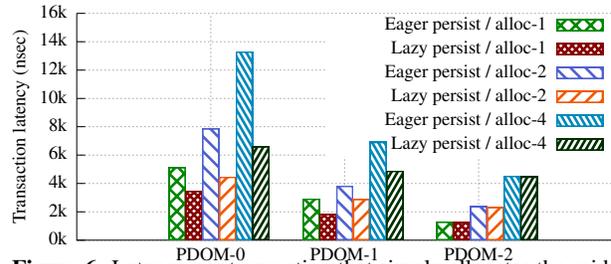}
\caption{Latency per transaction that simply allocates the said number
  of blocks (\textsf{Alloc-X}), each of a random size from 1 to 512
  bytes.}
\label{fig:alloc-latency}
\end{figure}

\autoref{fig:alloc-latency} shows memory allocation latency, comparing
the Eager Persist approach that uses persist barriers per
allocation/deallocation call, to our Lazy Persist approach that avoids
persist barriers altogether during allocation/deallocation calls.
There is no performance difference in PDOM-2, because the persist
barrier is a \nop. However, for PDOM-1, the optimization produces a
20--30\% latency improvement. In PDOM-0, the improvement grows to
30--100\%, because the persist barrier latency is much higher.

\subsection{Persistent Key-Value Store}
\label{s:perf-kvstore}

We implemented a persistent key-value (K-V) store from scratch using
our transactional interface.  The implementation served as a vehicle
to test the programmability limits of our transaction runtimes.
Our K-V store's central data structure is a concurrent, closed
addressed, hash table that uses chaining to resolve hash
collisions. The K-V store supports string-type keys and values, and
provides a simple get/put interface. Clients connect to the K-V store
via UNIX domain socket connections. The K-V store spawns a thread for
each connected client (we plan to extend our implementation to let
server threads handle multiple clients concurrently).

We started from an implementation that makes use of our transactional
API to perform all persistent data accesses.  We introduced persistent
types for all the persistent data structures.  This introduces
``wrapper'' objects for all the persistent objects hosted in our K-V
store.  The wrapper objects introduce a level of indirection and hence
overhead.  We also implemented a hand-optimized version of the K-V
store that avoids use of persistent wrapper types altogether.  Our
optimized version also aggressively bypasses the transactional
accessors wherever possible -- e.g. for read accesses, we can bypass
the \txnread\ accessors and fetch the data directly from the target
address in cases where the transaction has not yet written to the
address.  These optimizations can be trivially supported with the undo
and redo log runtimes.  COW, however, appears to have a fundamental
limitation -- it relies on the persistent wrappers to perform the
copy-on-write.  As a result, it is not possible to build such a
version of our K-V store using COW transactions.  This is a
significant limitation of the COW transaction interface.

\begin{figure}[t]
  \centering
  \input{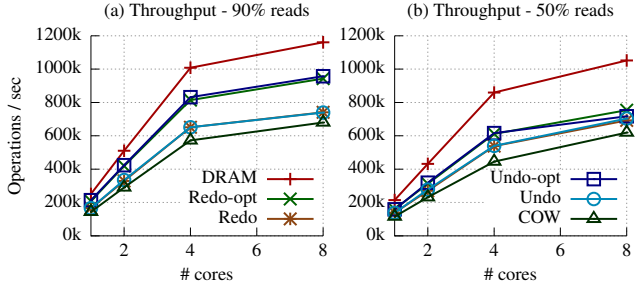}
\vspace{0.05in}
\caption{Persistent K-V store throughput. Suffixes ``-opt'' denote optimized
  versions.  Performance is measured assuming PDOM-2.
  Each utilized core serves one connected client, who issues 1 million operations
  according to the specified read-write ratio. The keys are randomly drawn from a
  space of 238,328 keys, and the K-V store is pre-populated by 1 million random
  \kvput{}s. Averages of 3 consecutive runs are reported. We omit error bars
  because of low variabilities in these benchmarks.}
\label{fig:ht}
\end{figure}

\autoref{fig:ht} shows performance of our various K-V store versions
for client (hence worker) thread count ranging from 1 to 8.  All
client threads are bound to 1 processor socket of the emulator, while
the worker threads are bound to the other socket.  Due to space
restrictions, we report just the PDOM-2 numbers.  COW experiences
cache locality overheads and performs worse than redo and undo
logging, which perform comparably.  We observed 12\% and 14\%
consecutive drops in performance for both redo and undo logging when
we ran the same experiments with PDOM-1 and PDOM-0 configurations.
COW also experiences similar performance drops.  Our optimizations of
bypassing accessors, persistent wrappers, and locality-friendly data
placement deliver performance improvements of as much as 27\% under
read-dominated workloads.  This difference comes down to 8\% for redo
logging and 2\% for undo logging under the 50\% reads workload.

Overall, we observe negligible difference between redo and undo
logging versions (both base and optimized versions).  The common case
hash table accesses (gets and puts) are extremely short transactions,
accessing a few cache lines, and updating even fewer (1--2) cache
lines transactionally (e.g. linking or unlinking a node from a hash
table bucket).  Additionally, they do not contain any read-after-write
accesses.  Furthermore, the workers receive requests from clients over
a TCP connection, which itself dominates the latency of client
operations.  This represents a possibly significant class of real
world workloads, where differences in these lower level abstraction
implementations may not matter to overall performance of the
application.

\subsection{SQLite}
\label{sec:sqlite}

SQLite~\cite{sqlite} is a popular light-weight relational database.
It hosts the entire database in a single file, with another file for
logging (rollback or write-ahead log) that is used to ensure ACID
semantics for its transactions.  SQLite can also be configured to an
``in-memory'' mode where the storage tier is completely removed.  The
database does not provide durability guarantees in that configuration.
We have extended this configuration to use our transactional API for
persistence.

SQLite stages all changes to the database at a page granularity in a
\emph{page cache}, and writes them out to the database file at the
commit phase. For in-memory databases, the page cache is still
populated, but it does not get persisted during the commit.  We built a
persistent version of in-memory SQLite by creating a ``region file''
based on our persistent region abstraction that is written to, and
persisted, during the commit phase using our transactional API.  The
dirty list is essentially applied in a single transaction, thus making
the transaction's effects durable in a failure atomic way.  Our
transaction effectively plays the role of rollback or write-ahead logs
in the stock SQLite configuration.

Since the region file is a single large object, COW transactions would
entail prohibitively high overheads, which is why we did not implement
SQLite with COW transactions.  We however tested our SQLite port with
both undo and redo log transactions, and compared them with the
in-memory configuration, and default SQLite databases whose files --
both the database file, and journal file -- were hosted in PMFS (we
tested memory-mapped files, which is a feature supported by SQLite,
but the results were identical to the default SQLite
configurations). Our modifications and tests were carried out on
SQLite 3.13.0.

\begin{figure}[t]
  \centering
  \input{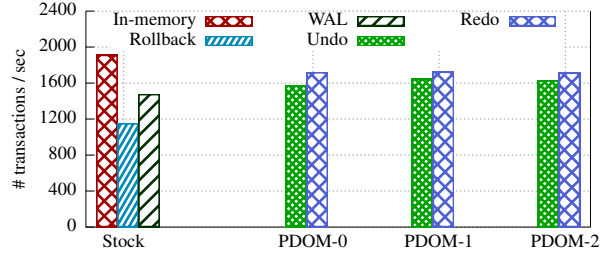}
\vspace{0.05in}
  \caption{TPC-C, as implemented by PyTpcc~\cite{tpcc}, results on
  SQLite. For each persistence domain, we present both undo and redo
  logging performance. In-memory databases and stock databases --
  default unmodified SQLite with files hosted in PMFS -- are presented
  for comparison.  In our transactional version, all transactions are
  write-only (no reads).}
\label{fig:sqlite}
\end{figure}

On all mentioned configurations, we ran the TPC-C~\cite{tpcc}
benchmark, as implemented by PyTpcc~\cite{tpcc}.  \autoref{fig:sqlite}
shows the results. For each configuration, we took the average of 3
runs.
As expected, the in-memory version has highest throughput
All our redo logging transactions have comparable throughput, just
10\% under the in-memory version.  Undo logging transactions in PDOM-1
and PDOM-2 configurations have 2-3\% lower throughput than the redo
logging ones; all this overhead is attributable to the checksum
overhead incurred for the page-granular transaction writes.
Furthermore, the PDOM-0 throughput of undo logging is 6\% lower than
the throughput of redo logging; this is where the 500 nanosecond
persist barrier latency shows up, approximately half of that overhead
is because of the persist barriers.  This performance nicely tracks
the performance we observed in the Array microbenchmark with slot size
of 64.  We expect such performance patterns to emerge in applications
that do lots of coarse grain writes.  The unmodified SQLite with a
rollback journal on PMFS are about 36\% to 50\% slower than the
in-memory database.

\subsection{Memcached}
\label{sec:memcached}

\begin{figure*}[t]
  \centering
  \input{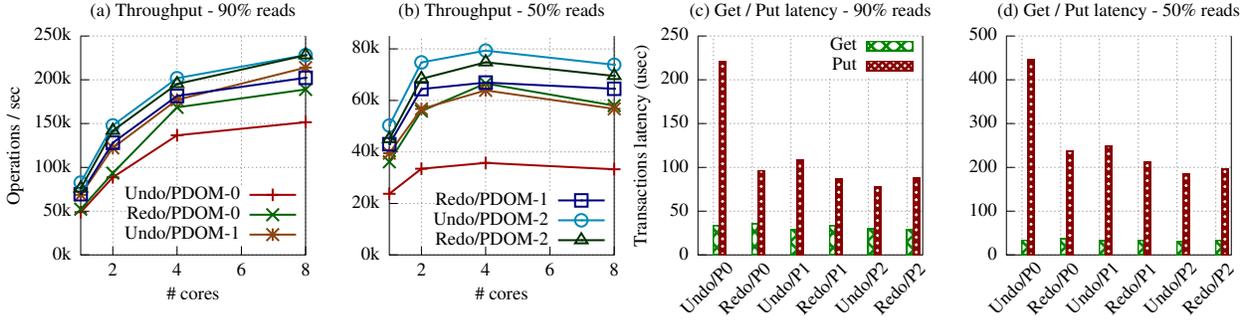}
\vspace{0.05in}
  \caption{\memcached\ scalability and latency results with 90/10\% and
  50/50\% \kvget{}/\kvput\ ratios.  Latency bar charts are for the
  8 thread runs. (P0 = PDOM-0, P1 = PDOM-1, P2 = PDOM-2.)}
\label{fig:memcached-g90-p10}
\end{figure*}

We have used our transactional API to ``persistify''
\memcached~\cite{memcached}, a widely used, high performance,
in-memory key-value store.  The motivation for building a persistent
version of \memcached\ is to accelerate the restart-warmup cycle after
a failure or shutdown, which can take several hours~\cite{goel14} in
some instances because \memcached\ is nonpersistent.  A persistent
\memcached\ can significantly accelerate the warmup time.  To that
end, the cache's state must be correctly persistified across failure
events such as power failures, kernel crashes, etc.  Failure atomic
updates are pivotal for this purpose.

We originally started the effort with the goal to simply port
\memcached{}'s central data structure, a concurrent, growable, closed
addressed hash table.  However, we quickly realized that we needed to
modify other parts of \memcached\ (LRU cache management, slab
allocator, lazy memory reclamation, etc.) to persist \memcached{}'s
entire internal state for warm restart.  As a result, this simple
effort evolved into a major port of \memcached\ to our transactional
API.  Transactions ended up encompassing fairly complex code paths
that led to some interesting scenarios such as: tiny critical sections
within transactions (supported using special capabilities such as
\emph{deferred operation execution} and \emph{deferred lock release},
provided in our library), semantically nonpersistent data located
within persistent data objects (supported using persistent generation
numbers), etc.

Additionally, \memcached\ itself is a copy-on-write based system, and
uses its own slab allocator for memory management.  This poses a
significant problem in using our COW-based transaction runtime to
perform updates since our runtime uses its own memory allocator for
copy-on-writes.  Furthermore, each key-value pair contains groups of
fields that are protected by different locks and can be modified
concurrently by multiple threads.  Our COW-based persistent objects
can be modified by just one thread at a time.  Overall, porting
\memcached\ to our COW-based transaction runtime seemed like a
significant enough restructuring of \memcached\ that we decided to not
do it.  This is another example of programmability challenges for
COW-based transaction runtimes.  We report results of our undo and
redo log based versions.

We evaluated \memcached\ using the \textsf{mutilate}
workload~\cite{atikoglu12}, fixing the number of client threads to 8.
We varied the number of \memcached\ worker threads from 1 to 8.
\autoref{fig:memcached-g90-p10} shows our persistent \memcached{}'s
performance with 90/10\% and 50/50\% \kvget{}/\kvput\ ratios.  First,
note that, for all thread counts, at 10\% \kvput{}s, the best
performing runtime \textsf{Undo/PDOM-2} has about 10--30\% lower
throughput than the original \memcached, whereas the same runtime has
about 45--60\% lower throughput than the original \memcached\ for 50\%
\kvput{}s.  This largely highlights the instrumentation and
bookkeeping overheads of our transaction runtimes for transactional
reads and writes.

Second, note that undo logging performs better than redo logging by
1-10\% in the 10\% \kvput\ tests for the PDOM-2 configuration.  This
difference is consistently closer to 10\% for the 50\% \kvput\ tests.
We gathered stastical data on read/write access patterns of
\memcached{}'s transactions.  What we found was quite interesting.
These transactions, particularly the \kvput\ transactions, appear to
have far more read-after-write instances (25-30\% of all reads done by
transactions) than we expected.

The \kvget\ transactions are short, averaging 25 reads, 3 instances of
which are read-after-writes, and 2 writes (to manage the LRU cache).
The \kvput\ operations break down into two distinct transactions: (i)
a transaction that allocates and initializes a new key-value pair
object, and (ii) a transaction to insert the new key-value pair, and
remove an old pair matching the key if one exists.  Both these
transactions contain 80--90 read accesses, approximately 25--30\% of
which are read-after-writes.  Furthermore, the transactions also
perform 24 writes on average.  A closer look at the source code
indicates that almost all read-after-write instances manifest when a
transaction first writes to one part of an object and then reads
another part of it (e.g. increments of multiple counters in an
object).  It appears that \memcached\ better matches our Array-RAW
microbenchmark's profile than Array's profile; the performance also
seems to nicely track that of Array-RAW (at about 25--30\%
read-after-write instances; see~\autoref{fig:ubench-latency}).
Similar to Array-RAW, the performance advantage undo logging has
diminishes as the persist barrier latency increases.

We also observe a difference in scalability patterns.  While
scalability for the PDOM-2 configuration is comparable between the
redo and undo logging implementations, undo logging scales worse than
redo logging for PDOM-1 and PDOM-0 configurations.  The explanation
appears in the corresponding latency bar chart in
\autoref{fig:memcached-g90-p10}(c), where we clearly see the latency
of \kvput\ operations go up significantly for undo logging (over 200
microseconds).  Our transactions end up inflating some of the critical
sections of \memcached\ (e.g. LRU cache management, slab allocator
management).  The higher latency of \kvput{}s leads to greater lock
hold intervals, which in turn hinders scalability.  With 50\%
\kvput{}s, for PDOM-0, the higher write latency (see
\autoref{fig:memcached-g90-p10}(d)) leads to significant slowdown in
undo logging transactions at all thread counts.

%% file: data/micro-new.tex
\begingroup
\scriptsize
  \makeatletter
  \providecommand\color[2][]{%
    \GenericError{(gnuplot) \space\space\space\@spaces}{%
      Package color not loaded in conjunction with
      terminal option `colourtext'%
    }{See the gnuplot documentation for explanation.%
    }{Either use 'blacktext' in gnuplot or load the package
      color.sty in LaTeX.}%
    \renewcommand\color[2][]{}%
  }%
  \providecommand\includegraphics[2][]{%
    \GenericError{(gnuplot) \space\space\space\@spaces}{%
      Package graphicx or graphics not loaded%
    }{See the gnuplot documentation for explanation.%
    }{The gnuplot epslatex terminal needs graphicx.sty or graphics.sty.}%
    \renewcommand\includegraphics[2][]{}%
  }%
  \providecommand\rotatebox[2]{#2}%
  \@ifundefined{ifGPcolor}{%
    \newif\ifGPcolor
    \GPcolortrue
  }{}%
  \@ifundefined{ifGPblacktext}{%
    \newif\ifGPblacktext
    \GPblacktextfalse
  }{}%
  \let\gplgaddtomacro\g@addto@macro
  \gdef\gplbacktext{}%
  \gdef\gplfronttext{}%
  \makeatother
  \ifGPblacktext
    \def\colorrgb#1{}%
    \def\colorgray#1{}%
  \else
    \ifGPcolor
      \def\colorrgb#1{\color[rgb]{#1}}%
      \def\colorgray#1{\color[gray]{#1}}%
      \expandafter\def\csname LTw\endcsname{\color{white}}%
      \expandafter\def\csname LTb\endcsname{\color{black}}%
      \expandafter\def\csname LTa\endcsname{\color{black}}%
      \expandafter\def\csname LT0\endcsname{\color[rgb]{1,0,0}}%
      \expandafter\def\csname LT1\endcsname{\color[rgb]{0,1,0}}%
      \expandafter\def\csname LT2\endcsname{\color[rgb]{0,0,1}}%
      \expandafter\def\csname LT3\endcsname{\color[rgb]{1,0,1}}%
      \expandafter\def\csname LT4\endcsname{\color[rgb]{0,1,1}}%
      \expandafter\def\csname LT5\endcsname{\color[rgb]{1,1,0}}%
      \expandafter\def\csname LT6\endcsname{\color[rgb]{0,0,0}}%
      \expandafter\def\csname LT7\endcsname{\color[rgb]{1,0.3,0}}%
      \expandafter\def\csname LT8\endcsname{\color[rgb]{0.5,0.5,0.5}}%
    \else
      \def\colorrgb#1{\color{black}}%
      \def\colorgray#1{\color[gray]{#1}}%
      \expandafter\def\csname LTw\endcsname{\color{white}}%
      \expandafter\def\csname LTb\endcsname{\color{black}}%
      \expandafter\def\csname LTa\endcsname{\color{black}}%
      \expandafter\def\csname LT0\endcsname{\color{black}}%
      \expandafter\def\csname LT1\endcsname{\color{black}}%
      \expandafter\def\csname LT2\endcsname{\color{black}}%
      \expandafter\def\csname LT3\endcsname{\color{black}}%
      \expandafter\def\csname LT4\endcsname{\color{black}}%
      \expandafter\def\csname LT5\endcsname{\color{black}}%
      \expandafter\def\csname LT6\endcsname{\color{black}}%
      \expandafter\def\csname LT7\endcsname{\color{black}}%
      \expandafter\def\csname LT8\endcsname{\color{black}}%
    \fi
  \fi
    \setlength{\unitlength}{0.0500bp}%
    \ifx\gptboxheight\undefined%
      \newlength{\gptboxheight}%
      \newlength{\gptboxwidth}%
      \newsavebox{\gptboxtext}%
    \fi%
    \setlength{\fboxrule}{0.5pt}%
    \setlength{\fboxsep}{1pt}%
\begin{picture}(9792.00,6912.00)%
    \gplgaddtomacro\gplbacktext{%
      \csname LTb\endcsname%
      \put(907,5234){\makebox(0,0)[r]{\strut{} 2k}}%
      \csname LTb\endcsname%
      \put(907,5480){\makebox(0,0)[r]{\strut{} 4k}}%
      \csname LTb\endcsname%
      \put(907,5727){\makebox(0,0)[r]{\strut{} 8k}}%
      \csname LTb\endcsname%
      \put(907,5973){\makebox(0,0)[r]{\strut{}16k}}%
      \csname LTb\endcsname%
      \put(907,6219){\makebox(0,0)[r]{\strut{}32k}}%
      \csname LTb\endcsname%
      \put(979,5012){\makebox(0,0){\strut{}$0$}}%
      \csname LTb\endcsname%
      \put(1358,5012){\makebox(0,0){\strut{}$20$}}%
      \csname LTb\endcsname%
      \put(1736,5012){\makebox(0,0){\strut{}$40$}}%
      \csname LTb\endcsname%
      \put(2115,5012){\makebox(0,0){\strut{}$60$}}%
      \csname LTb\endcsname%
      \put(2493,5012){\makebox(0,0){\strut{}$80$}}%
      \csname LTb\endcsname%
      \put(2872,5012){\makebox(0,0){\strut{}$100$}}%
    }%
    \gplgaddtomacro\gplfronttext{%
      \csname LTb\endcsname%
      \put(559,5675){\rotatebox{-270}{\makebox(0,0){\strut{}Avg. Latency (nsecs)}}}%
      \put(1925,4832){\makebox(0,0){\strut{}Write Percentage}}%
      \put(1925,6399){\makebox(0,0){\strut{}(a) Array Slot Size (4), PDOM-0}}%
      \csname LTb\endcsname%
      \put(2377,5358){\makebox(0,0)[r]{\strut{}COW}}%
    }%
    \gplgaddtomacro\gplbacktext{%
      \csname LTb\endcsname%
      \put(3877,5234){\makebox(0,0)[r]{\strut{} 2k}}%
      \csname LTb\endcsname%
      \put(3877,5480){\makebox(0,0)[r]{\strut{} 4k}}%
      \csname LTb\endcsname%
      \put(3877,5727){\makebox(0,0)[r]{\strut{} 8k}}%
      \csname LTb\endcsname%
      \put(3877,5973){\makebox(0,0)[r]{\strut{}16k}}%
      \csname LTb\endcsname%
      \put(3877,6219){\makebox(0,0)[r]{\strut{}32k}}%
      \csname LTb\endcsname%
      \put(3949,5012){\makebox(0,0){\strut{}$0$}}%
      \csname LTb\endcsname%
      \put(4327,5012){\makebox(0,0){\strut{}$20$}}%
      \csname LTb\endcsname%
      \put(4706,5012){\makebox(0,0){\strut{}$40$}}%
      \csname LTb\endcsname%
      \put(5084,5012){\makebox(0,0){\strut{}$60$}}%
      \csname LTb\endcsname%
      \put(5463,5012){\makebox(0,0){\strut{}$80$}}%
      \csname LTb\endcsname%
      \put(5841,5012){\makebox(0,0){\strut{}$100$}}%
    }%
    \gplgaddtomacro\gplfronttext{%
      \csname LTb\endcsname%
      \put(3529,5675){\rotatebox{-270}{\makebox(0,0){\strut{}Avg. Latency (nsecs)}}}%
      \put(4895,4832){\makebox(0,0){\strut{}Write Percentage}}%
      \put(4895,6399){\makebox(0,0){\strut{}(b) Array Slot Size (4), PDOM-1}}%
      \csname LTb\endcsname%
      \put(5346,5358){\makebox(0,0)[r]{\strut{}Undo}}%
    }%
    \gplgaddtomacro\gplbacktext{%
      \csname LTb\endcsname%
      \put(6847,5234){\makebox(0,0)[r]{\strut{} 2k}}%
      \csname LTb\endcsname%
      \put(6847,5480){\makebox(0,0)[r]{\strut{} 4k}}%
      \csname LTb\endcsname%
      \put(6847,5727){\makebox(0,0)[r]{\strut{} 8k}}%
      \csname LTb\endcsname%
      \put(6847,5973){\makebox(0,0)[r]{\strut{}16k}}%
      \csname LTb\endcsname%
      \put(6847,6219){\makebox(0,0)[r]{\strut{}32k}}%
      \csname LTb\endcsname%
      \put(6919,5012){\makebox(0,0){\strut{}$0$}}%
      \csname LTb\endcsname%
      \put(7297,5012){\makebox(0,0){\strut{}$20$}}%
      \csname LTb\endcsname%
      \put(7676,5012){\makebox(0,0){\strut{}$40$}}%
      \csname LTb\endcsname%
      \put(8054,5012){\makebox(0,0){\strut{}$60$}}%
      \csname LTb\endcsname%
      \put(8433,5012){\makebox(0,0){\strut{}$80$}}%
      \csname LTb\endcsname%
      \put(8811,5012){\makebox(0,0){\strut{}$100$}}%
    }%
    \gplgaddtomacro\gplfronttext{%
      \csname LTb\endcsname%
      \put(6499,5675){\rotatebox{-270}{\makebox(0,0){\strut{}Avg. Latency (nsecs)}}}%
      \put(7865,4832){\makebox(0,0){\strut{}Write Percentage}}%
      \put(7865,6399){\makebox(0,0){\strut{}(c) Array Slot Size (4), PDOM-2}}%
      \csname LTb\endcsname%
      \put(8316,5358){\makebox(0,0)[r]{\strut{}Redo}}%
    }%
    \gplgaddtomacro\gplbacktext{%
      \csname LTb\endcsname%
      \put(907,3421){\makebox(0,0)[r]{\strut{}   4k}}%
      \csname LTb\endcsname%
      \put(907,3557){\makebox(0,0)[r]{\strut{}   8k}}%
      \csname LTb\endcsname%
      \put(907,3693){\makebox(0,0)[r]{\strut{}  16k}}%
      \csname LTb\endcsname%
      \put(907,3829){\makebox(0,0)[r]{\strut{}  32k}}%
      \csname LTb\endcsname%
      \put(907,3965){\makebox(0,0)[r]{\strut{}  64k}}%
      \csname LTb\endcsname%
      \put(907,4101){\makebox(0,0)[r]{\strut{} 128k}}%
      \csname LTb\endcsname%
      \put(907,4237){\makebox(0,0)[r]{\strut{} 256k}}%
      \csname LTb\endcsname%
      \put(907,4373){\makebox(0,0)[r]{\strut{} 512k}}%
      \csname LTb\endcsname%
      \put(907,4509){\makebox(0,0)[r]{\strut{}1024k}}%
      \csname LTb\endcsname%
      \put(979,3301){\makebox(0,0){\strut{}$0$}}%
      \csname LTb\endcsname%
      \put(1358,3301){\makebox(0,0){\strut{}$20$}}%
      \csname LTb\endcsname%
      \put(1736,3301){\makebox(0,0){\strut{}$40$}}%
      \csname LTb\endcsname%
      \put(2115,3301){\makebox(0,0){\strut{}$60$}}%
      \csname LTb\endcsname%
      \put(2493,3301){\makebox(0,0){\strut{}$80$}}%
      \csname LTb\endcsname%
      \put(2872,3301){\makebox(0,0){\strut{}$100$}}%
    }%
    \gplgaddtomacro\gplfronttext{%
      \csname LTb\endcsname%
      \put(415,3965){\rotatebox{-270}{\makebox(0,0){\strut{}Avg. Latency (nsecs)}}}%
      \put(1925,3121){\makebox(0,0){\strut{}Write Percentage}}%
      \put(1925,4689){\makebox(0,0){\strut{}(d) Array Slot Size (64), PDOM-0}}%
    }%
    \gplgaddtomacro\gplbacktext{%
      \csname LTb\endcsname%
      \put(3877,3421){\makebox(0,0)[r]{\strut{}   4k}}%
      \csname LTb\endcsname%
      \put(3877,3557){\makebox(0,0)[r]{\strut{}   8k}}%
      \csname LTb\endcsname%
      \put(3877,3693){\makebox(0,0)[r]{\strut{}  16k}}%
      \csname LTb\endcsname%
      \put(3877,3829){\makebox(0,0)[r]{\strut{}  32k}}%
      \csname LTb\endcsname%
      \put(3877,3965){\makebox(0,0)[r]{\strut{}  64k}}%
      \csname LTb\endcsname%
      \put(3877,4101){\makebox(0,0)[r]{\strut{} 128k}}%
      \csname LTb\endcsname%
      \put(3877,4237){\makebox(0,0)[r]{\strut{} 256k}}%
      \csname LTb\endcsname%
      \put(3877,4373){\makebox(0,0)[r]{\strut{} 512k}}%
      \csname LTb\endcsname%
      \put(3877,4509){\makebox(0,0)[r]{\strut{}1024k}}%
      \csname LTb\endcsname%
      \put(3949,3301){\makebox(0,0){\strut{}$0$}}%
      \csname LTb\endcsname%
      \put(4327,3301){\makebox(0,0){\strut{}$20$}}%
      \csname LTb\endcsname%
      \put(4706,3301){\makebox(0,0){\strut{}$40$}}%
      \csname LTb\endcsname%
      \put(5084,3301){\makebox(0,0){\strut{}$60$}}%
      \csname LTb\endcsname%
      \put(5463,3301){\makebox(0,0){\strut{}$80$}}%
      \csname LTb\endcsname%
      \put(5841,3301){\makebox(0,0){\strut{}$100$}}%
    }%
    \gplgaddtomacro\gplfronttext{%
      \csname LTb\endcsname%
      \put(3385,3965){\rotatebox{-270}{\makebox(0,0){\strut{}Avg. Latency (nsecs)}}}%
      \put(4895,3121){\makebox(0,0){\strut{}Write Percentage}}%
      \put(4895,4689){\makebox(0,0){\strut{}(e) Array Slot Size (64), PDOM-1}}%
    }%
    \gplgaddtomacro\gplbacktext{%
      \csname LTb\endcsname%
      \put(6847,3421){\makebox(0,0)[r]{\strut{}   4k}}%
      \csname LTb\endcsname%
      \put(6847,3557){\makebox(0,0)[r]{\strut{}   8k}}%
      \csname LTb\endcsname%
      \put(6847,3693){\makebox(0,0)[r]{\strut{}  16k}}%
      \csname LTb\endcsname%
      \put(6847,3829){\makebox(0,0)[r]{\strut{}  32k}}%
      \csname LTb\endcsname%
      \put(6847,3965){\makebox(0,0)[r]{\strut{}  64k}}%
      \csname LTb\endcsname%
      \put(6847,4101){\makebox(0,0)[r]{\strut{} 128k}}%
      \csname LTb\endcsname%
      \put(6847,4237){\makebox(0,0)[r]{\strut{} 256k}}%
      \csname LTb\endcsname%
      \put(6847,4373){\makebox(0,0)[r]{\strut{} 512k}}%
      \csname LTb\endcsname%
      \put(6847,4509){\makebox(0,0)[r]{\strut{}1024k}}%
      \csname LTb\endcsname%
      \put(6919,3301){\makebox(0,0){\strut{}$0$}}%
      \csname LTb\endcsname%
      \put(7297,3301){\makebox(0,0){\strut{}$20$}}%
      \csname LTb\endcsname%
      \put(7676,3301){\makebox(0,0){\strut{}$40$}}%
      \csname LTb\endcsname%
      \put(8054,3301){\makebox(0,0){\strut{}$60$}}%
      \csname LTb\endcsname%
      \put(8433,3301){\makebox(0,0){\strut{}$80$}}%
      \csname LTb\endcsname%
      \put(8811,3301){\makebox(0,0){\strut{}$100$}}%
    }%
    \gplgaddtomacro\gplfronttext{%
      \csname LTb\endcsname%
      \put(6355,3965){\rotatebox{-270}{\makebox(0,0){\strut{}Avg. Latency (nsecs)}}}%
      \put(7865,3121){\makebox(0,0){\strut{}Write Percentage}}%
      \put(7865,4689){\makebox(0,0){\strut{}(f) Array Slot Size (64), PDOM-2}}%
    }%
    \gplgaddtomacro\gplbacktext{%
      \csname LTb\endcsname%
      \put(907,1710){\makebox(0,0)[r]{\strut{}  2k}}%
      \csname LTb\endcsname%
      \put(907,1891){\makebox(0,0)[r]{\strut{}  4k}}%
      \csname LTb\endcsname%
      \put(907,2073){\makebox(0,0)[r]{\strut{}  8k}}%
      \csname LTb\endcsname%
      \put(907,2254){\makebox(0,0)[r]{\strut{} 16k}}%
      \csname LTb\endcsname%
      \put(907,2435){\makebox(0,0)[r]{\strut{} 32k}}%
      \csname LTb\endcsname%
      \put(907,2617){\makebox(0,0)[r]{\strut{} 64k}}%
      \csname LTb\endcsname%
      \put(907,2798){\makebox(0,0)[r]{\strut{}128k}}%
      \csname LTb\endcsname%
      \put(979,1590){\makebox(0,0){\strut{}$0$}}%
      \csname LTb\endcsname%
      \put(1358,1590){\makebox(0,0){\strut{}$20$}}%
      \csname LTb\endcsname%
      \put(1736,1590){\makebox(0,0){\strut{}$40$}}%
      \csname LTb\endcsname%
      \put(2115,1590){\makebox(0,0){\strut{}$60$}}%
      \csname LTb\endcsname%
      \put(2493,1590){\makebox(0,0){\strut{}$80$}}%
      \csname LTb\endcsname%
      \put(2872,1590){\makebox(0,0){\strut{}$100$}}%
    }%
    \gplgaddtomacro\gplfronttext{%
      \csname LTb\endcsname%
      \put(487,2254){\rotatebox{-270}{\makebox(0,0){\strut{}Avg. Latency (nsecs)}}}%
      \put(1925,1410){\makebox(0,0){\strut{}Write Percentage}}%
      \put(1925,2978){\makebox(0,0){\strut{}(g) Array-RAW Slot Size (4), PDOM-0}}%
    }%
    \gplgaddtomacro\gplbacktext{%
      \csname LTb\endcsname%
      \put(3877,1710){\makebox(0,0)[r]{\strut{}  2k}}%
      \csname LTb\endcsname%
      \put(3877,1928){\makebox(0,0)[r]{\strut{}  4k}}%
      \csname LTb\endcsname%
      \put(3877,2145){\makebox(0,0)[r]{\strut{}  8k}}%
      \csname LTb\endcsname%
      \put(3877,2363){\makebox(0,0)[r]{\strut{} 16k}}%
      \csname LTb\endcsname%
      \put(3877,2580){\makebox(0,0)[r]{\strut{} 32k}}%
      \csname LTb\endcsname%
      \put(3877,2798){\makebox(0,0)[r]{\strut{} 64k}}%
      \csname LTb\endcsname%
      \put(3949,1590){\makebox(0,0){\strut{}$0$}}%
      \csname LTb\endcsname%
      \put(4327,1590){\makebox(0,0){\strut{}$20$}}%
      \csname LTb\endcsname%
      \put(4706,1590){\makebox(0,0){\strut{}$40$}}%
      \csname LTb\endcsname%
      \put(5084,1590){\makebox(0,0){\strut{}$60$}}%
      \csname LTb\endcsname%
      \put(5463,1590){\makebox(0,0){\strut{}$80$}}%
      \csname LTb\endcsname%
      \put(5841,1590){\makebox(0,0){\strut{}$100$}}%
    }%
    \gplgaddtomacro\gplfronttext{%
      \csname LTb\endcsname%
      \put(3457,2254){\rotatebox{-270}{\makebox(0,0){\strut{}Avg. Latency (nsecs)}}}%
      \put(4895,1410){\makebox(0,0){\strut{}Write Percentage}}%
      \put(4895,2978){\makebox(0,0){\strut{}(h) Array-RAW Slot Size (4), PDOM-1}}%
    }%
    \gplgaddtomacro\gplbacktext{%
      \csname LTb\endcsname%
      \put(6847,1710){\makebox(0,0)[r]{\strut{}  2k}}%
      \csname LTb\endcsname%
      \put(6847,1928){\makebox(0,0)[r]{\strut{}  4k}}%
      \csname LTb\endcsname%
      \put(6847,2145){\makebox(0,0)[r]{\strut{}  8k}}%
      \csname LTb\endcsname%
      \put(6847,2363){\makebox(0,0)[r]{\strut{} 16k}}%
      \csname LTb\endcsname%
      \put(6847,2580){\makebox(0,0)[r]{\strut{} 32k}}%
      \csname LTb\endcsname%
      \put(6847,2798){\makebox(0,0)[r]{\strut{} 64k}}%
      \csname LTb\endcsname%
      \put(6919,1590){\makebox(0,0){\strut{}$0$}}%
      \csname LTb\endcsname%
      \put(7297,1590){\makebox(0,0){\strut{}$20$}}%
      \csname LTb\endcsname%
      \put(7676,1590){\makebox(0,0){\strut{}$40$}}%
      \csname LTb\endcsname%
      \put(8054,1590){\makebox(0,0){\strut{}$60$}}%
      \csname LTb\endcsname%
      \put(8433,1590){\makebox(0,0){\strut{}$80$}}%
      \csname LTb\endcsname%
      \put(8811,1590){\makebox(0,0){\strut{}$100$}}%
    }%
    \gplgaddtomacro\gplfronttext{%
      \csname LTb\endcsname%
      \put(6427,2254){\rotatebox{-270}{\makebox(0,0){\strut{}Avg. Latency (nsecs)}}}%
      \put(7865,1410){\makebox(0,0){\strut{}Write Percentage}}%
      \put(7865,2978){\makebox(0,0){\strut{}(i) Array-RAW Slot Size (4), PDOM-2}}%
    }%
    \gplgaddtomacro\gplbacktext{%
      \csname LTb\endcsname%
      \put(907,61){\makebox(0,0)[r]{\strut{}  32k}}%
      \csname LTb\endcsname%
      \put(907,232){\makebox(0,0)[r]{\strut{}  64k}}%
      \csname LTb\endcsname%
      \put(907,403){\makebox(0,0)[r]{\strut{} 128k}}%
      \csname LTb\endcsname%
      \put(907,574){\makebox(0,0)[r]{\strut{} 256k}}%
      \csname LTb\endcsname%
      \put(907,746){\makebox(0,0)[r]{\strut{} 512k}}%
      \csname LTb\endcsname%
      \put(907,917){\makebox(0,0)[r]{\strut{}1024k}}%
      \csname LTb\endcsname%
      \put(907,1088){\makebox(0,0)[r]{\strut{}2048k}}%
      \csname LTb\endcsname%
      \put(979,-120){\makebox(0,0){\strut{}$0$}}%
      \csname LTb\endcsname%
      \put(1358,-120){\makebox(0,0){\strut{}$20$}}%
      \csname LTb\endcsname%
      \put(1736,-120){\makebox(0,0){\strut{}$40$}}%
      \csname LTb\endcsname%
      \put(2115,-120){\makebox(0,0){\strut{}$60$}}%
      \csname LTb\endcsname%
      \put(2493,-120){\makebox(0,0){\strut{}$80$}}%
      \csname LTb\endcsname%
      \put(2872,-120){\makebox(0,0){\strut{}$100$}}%
    }%
    \gplgaddtomacro\gplfronttext{%
      \csname LTb\endcsname%
      \put(415,544){\rotatebox{-270}{\makebox(0,0){\strut{}Avg. Latency (nsecs)}}}%
      \put(1925,-300){\makebox(0,0){\strut{}Write Percentage}}%
      \put(1925,1268){\makebox(0,0){\strut{}(j) Array-RAW Slot Size (64), PDOM-0}}%
    }%
    \gplgaddtomacro\gplbacktext{%
      \csname LTb\endcsname%
      \put(3877,72){\makebox(0,0)[r]{\strut{}  32k}}%
      \csname LTb\endcsname%
      \put(3877,275){\makebox(0,0)[r]{\strut{}  64k}}%
      \csname LTb\endcsname%
      \put(3877,479){\makebox(0,0)[r]{\strut{} 128k}}%
      \csname LTb\endcsname%
      \put(3877,682){\makebox(0,0)[r]{\strut{} 256k}}%
      \csname LTb\endcsname%
      \put(3877,885){\makebox(0,0)[r]{\strut{} 512k}}%
      \csname LTb\endcsname%
      \put(3877,1088){\makebox(0,0)[r]{\strut{}1024k}}%
      \csname LTb\endcsname%
      \put(3949,-120){\makebox(0,0){\strut{}$0$}}%
      \csname LTb\endcsname%
      \put(4327,-120){\makebox(0,0){\strut{}$20$}}%
      \csname LTb\endcsname%
      \put(4706,-120){\makebox(0,0){\strut{}$40$}}%
      \csname LTb\endcsname%
      \put(5084,-120){\makebox(0,0){\strut{}$60$}}%
      \csname LTb\endcsname%
      \put(5463,-120){\makebox(0,0){\strut{}$80$}}%
      \csname LTb\endcsname%
      \put(5841,-120){\makebox(0,0){\strut{}$100$}}%
    }%
    \gplgaddtomacro\gplfronttext{%
      \csname LTb\endcsname%
      \put(3385,544){\rotatebox{-270}{\makebox(0,0){\strut{}Avg. Latency (nsecs)}}}%
      \put(4895,-300){\makebox(0,0){\strut{}Write Percentage}}%
      \put(4895,1268){\makebox(0,0){\strut{}(k) Array-RAW Slot Size (64), PDOM-1}}%
    }%
    \gplgaddtomacro\gplbacktext{%
      \csname LTb\endcsname%
      \put(6847,72){\makebox(0,0)[r]{\strut{}  32k}}%
      \csname LTb\endcsname%
      \put(6847,275){\makebox(0,0)[r]{\strut{}  64k}}%
      \csname LTb\endcsname%
      \put(6847,479){\makebox(0,0)[r]{\strut{} 128k}}%
      \csname LTb\endcsname%
      \put(6847,682){\makebox(0,0)[r]{\strut{} 256k}}%
      \csname LTb\endcsname%
      \put(6847,885){\makebox(0,0)[r]{\strut{} 512k}}%
      \csname LTb\endcsname%
      \put(6847,1088){\makebox(0,0)[r]{\strut{}1024k}}%
      \csname LTb\endcsname%
      \put(6919,-120){\makebox(0,0){\strut{}$0$}}%
      \csname LTb\endcsname%
      \put(7297,-120){\makebox(0,0){\strut{}$20$}}%
      \csname LTb\endcsname%
      \put(7676,-120){\makebox(0,0){\strut{}$40$}}%
      \csname LTb\endcsname%
      \put(8054,-120){\makebox(0,0){\strut{}$60$}}%
      \csname LTb\endcsname%
      \put(8433,-120){\makebox(0,0){\strut{}$80$}}%
      \csname LTb\endcsname%
      \put(8811,-120){\makebox(0,0){\strut{}$100$}}%
    }%
    \gplgaddtomacro\gplfronttext{%
      \csname LTb\endcsname%
      \put(6355,544){\rotatebox{-270}{\makebox(0,0){\strut{}Avg. Latency (nsecs)}}}%
      \put(7865,-300){\makebox(0,0){\strut{}Write Percentage}}%
      \put(7865,1268){\makebox(0,0){\strut{}(l) Array-RAW Slot Size (64), PDOM-2}}%
    }%
    \gplbacktext
    \put(-1000,-1000){\includegraphics{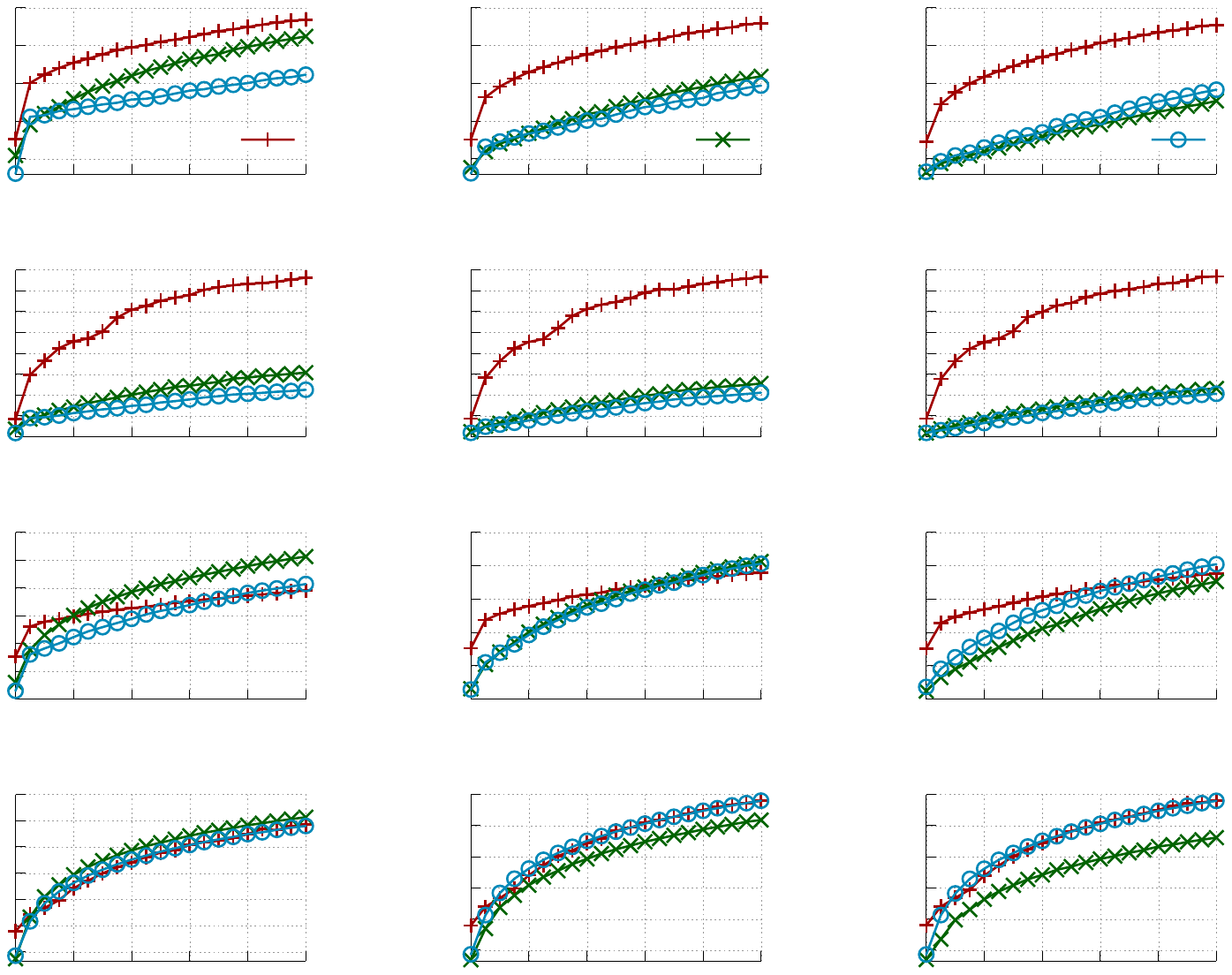}}%
    \gplfronttext
  \end{picture}%
\endgroup

%% file: data/alloc-latency.tex
\begingroup
\scriptsize
  \makeatletter
  \providecommand\color[2][]{%
    \GenericError{(gnuplot) \space\space\space\@spaces}{%
      Package color not loaded in conjunction with
      terminal option `colourtext'%
    }{See the gnuplot documentation for explanation.%
    }{Either use 'blacktext' in gnuplot or load the package
      color.sty in LaTeX.}%
    \renewcommand\color[2][]{}%
  }%
  \providecommand\includegraphics[2][]{%
    \GenericError{(gnuplot) \space\space\space\@spaces}{%
      Package graphicx or graphics not loaded%
    }{See the gnuplot documentation for explanation.%
    }{The gnuplot epslatex terminal needs graphicx.sty or graphics.sty.}%
    \renewcommand\includegraphics[2][]{}%
  }%
  \providecommand\rotatebox[2]{#2}%
  \@ifundefined{ifGPcolor}{%
    \newif\ifGPcolor
    \GPcolortrue
  }{}%
  \@ifundefined{ifGPblacktext}{%
    \newif\ifGPblacktext
    \GPblacktextfalse
  }{}%
  \let\gplgaddtomacro\g@addto@macro
  \gdef\gplbacktext{}%
  \gdef\gplfronttext{}%
  \makeatother
  \ifGPblacktext
    \def\colorrgb#1{}%
    \def\colorgray#1{}%
  \else
    \ifGPcolor
      \def\colorrgb#1{\color[rgb]{#1}}%
      \def\colorgray#1{\color[gray]{#1}}%
      \expandafter\def\csname LTw\endcsname{\color{white}}%
      \expandafter\def\csname LTb\endcsname{\color{black}}%
      \expandafter\def\csname LTa\endcsname{\color{black}}%
      \expandafter\def\csname LT0\endcsname{\color[rgb]{1,0,0}}%
      \expandafter\def\csname LT1\endcsname{\color[rgb]{0,1,0}}%
      \expandafter\def\csname LT2\endcsname{\color[rgb]{0,0,1}}%
      \expandafter\def\csname LT3\endcsname{\color[rgb]{1,0,1}}%
      \expandafter\def\csname LT4\endcsname{\color[rgb]{0,1,1}}%
      \expandafter\def\csname LT5\endcsname{\color[rgb]{1,1,0}}%
      \expandafter\def\csname LT6\endcsname{\color[rgb]{0,0,0}}%
      \expandafter\def\csname LT7\endcsname{\color[rgb]{1,0.3,0}}%
      \expandafter\def\csname LT8\endcsname{\color[rgb]{0.5,0.5,0.5}}%
    \else
      \def\colorrgb#1{\color{black}}%
      \def\colorgray#1{\color[gray]{#1}}%
      \expandafter\def\csname LTw\endcsname{\color{white}}%
      \expandafter\def\csname LTb\endcsname{\color{black}}%
      \expandafter\def\csname LTa\endcsname{\color{black}}%
      \expandafter\def\csname LT0\endcsname{\color{black}}%
      \expandafter\def\csname LT1\endcsname{\color{black}}%
      \expandafter\def\csname LT2\endcsname{\color{black}}%
      \expandafter\def\csname LT3\endcsname{\color{black}}%
      \expandafter\def\csname LT4\endcsname{\color{black}}%
      \expandafter\def\csname LT5\endcsname{\color{black}}%
      \expandafter\def\csname LT6\endcsname{\color{black}}%
      \expandafter\def\csname LT7\endcsname{\color{black}}%
      \expandafter\def\csname LT8\endcsname{\color{black}}%
    \fi
  \fi
    \setlength{\unitlength}{0.0500bp}%
    \ifx\gptboxheight\undefined%
      \newlength{\gptboxheight}%
      \newlength{\gptboxwidth}%
      \newsavebox{\gptboxtext}%
    \fi%
    \setlength{\fboxrule}{0.5pt}%
    \setlength{\fboxsep}{1pt}%
\begin{picture}(4896.00,2014.00)%
    \gplgaddtomacro\gplbacktext{%
      \csname LTb\endcsname%
      \put(417,100){\makebox(0,0)[r]{\strut{} 0k}}%
      \csname LTb\endcsname%
      \put(417,314){\makebox(0,0)[r]{\strut{} 2k}}%
      \csname LTb\endcsname%
      \put(417,528){\makebox(0,0)[r]{\strut{} 4k}}%
      \csname LTb\endcsname%
      \put(417,742){\makebox(0,0)[r]{\strut{} 6k}}%
      \csname LTb\endcsname%
      \put(417,956){\makebox(0,0)[r]{\strut{} 8k}}%
      \csname LTb\endcsname%
      \put(417,1169){\makebox(0,0)[r]{\strut{}10k}}%
      \csname LTb\endcsname%
      \put(417,1383){\makebox(0,0)[r]{\strut{}12k}}%
      \csname LTb\endcsname%
      \put(417,1597){\makebox(0,0)[r]{\strut{}14k}}%
      \csname LTb\endcsname%
      \put(417,1811){\makebox(0,0)[r]{\strut{}16k}}%
      \csname LTb\endcsname%
      \put(1499,-20){\makebox(0,0){\strut{}PDOM-0}}%
      \csname LTb\endcsname%
      \put(2508,-20){\makebox(0,0){\strut{}PDOM-1}}%
      \csname LTb\endcsname%
      \put(3518,-20){\makebox(0,0){\strut{}PDOM-2}}%
    }%
    \gplgaddtomacro\gplfronttext{%
      \csname LTb\endcsname%
      \put(69,955){\rotatebox{-270}{\makebox(0,0){\strut{}Transaction latency (nsec)}}}%
      \csname LTb\endcsname%
      \put(4032,1721){\makebox(0,0)[r]{\strut{}Eager persist / alloc-1}}%
      \csname LTb\endcsname%
      \put(4032,1541){\makebox(0,0)[r]{\strut{}Lazy persist / alloc-1}}%
      \csname LTb\endcsname%
      \put(4032,1361){\makebox(0,0)[r]{\strut{}Eager persist / alloc-2}}%
      \csname LTb\endcsname%
      \put(4032,1181){\makebox(0,0)[r]{\strut{}Lazy persist / alloc-2}}%
      \csname LTb\endcsname%
      \put(4032,1001){\makebox(0,0)[r]{\strut{}Eager persist / alloc-4}}%
      \csname LTb\endcsname%
      \put(4032,821){\makebox(0,0)[r]{\strut{}Lazy persist / alloc-4}}%
    }%
    \gplbacktext
    \put(-1000,-1000){\includegraphics{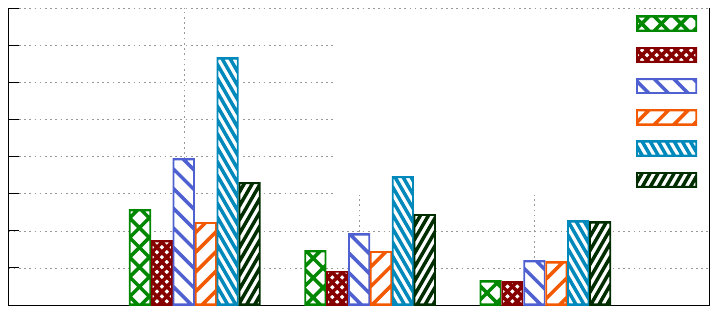}}%
    \gplfronttext
  \end{picture}%
\endgroup

%% file: data/ht.tex
\begingroup
\scriptsize
  \makeatletter
  \providecommand\color[2][]{%
    \GenericError{(gnuplot) \space\space\space\@spaces}{%
      Package color not loaded in conjunction with
      terminal option `colourtext'%
    }{See the gnuplot documentation for explanation.%
    }{Either use 'blacktext' in gnuplot or load the package
      color.sty in LaTeX.}%
    \renewcommand\color[2][]{}%
  }%
  \providecommand\includegraphics[2][]{%
    \GenericError{(gnuplot) \space\space\space\@spaces}{%
      Package graphicx or graphics not loaded%
    }{See the gnuplot documentation for explanation.%
    }{The gnuplot epslatex terminal needs graphicx.sty or graphics.sty.}%
    \renewcommand\includegraphics[2][]{}%
  }%
  \providecommand\rotatebox[2]{#2}%
  \@ifundefined{ifGPcolor}{%
    \newif\ifGPcolor
    \GPcolortrue
  }{}%
  \@ifundefined{ifGPblacktext}{%
    \newif\ifGPblacktext
    \GPblacktextfalse
  }{}%
  \let\gplgaddtomacro\g@addto@macro
  \gdef\gplbacktext{}%
  \gdef\gplfronttext{}%
  \makeatother
  \ifGPblacktext
    \def\colorrgb#1{}%
    \def\colorgray#1{}%
  \else
    \ifGPcolor
      \def\colorrgb#1{\color[rgb]{#1}}%
      \def\colorgray#1{\color[gray]{#1}}%
      \expandafter\def\csname LTw\endcsname{\color{white}}%
      \expandafter\def\csname LTb\endcsname{\color{black}}%
      \expandafter\def\csname LTa\endcsname{\color{black}}%
      \expandafter\def\csname LT0\endcsname{\color[rgb]{1,0,0}}%
      \expandafter\def\csname LT1\endcsname{\color[rgb]{0,1,0}}%
      \expandafter\def\csname LT2\endcsname{\color[rgb]{0,0,1}}%
      \expandafter\def\csname LT3\endcsname{\color[rgb]{1,0,1}}%
      \expandafter\def\csname LT4\endcsname{\color[rgb]{0,1,1}}%
      \expandafter\def\csname LT5\endcsname{\color[rgb]{1,1,0}}%
      \expandafter\def\csname LT6\endcsname{\color[rgb]{0,0,0}}%
      \expandafter\def\csname LT7\endcsname{\color[rgb]{1,0.3,0}}%
      \expandafter\def\csname LT8\endcsname{\color[rgb]{0.5,0.5,0.5}}%
    \else
      \def\colorrgb#1{\color{black}}%
      \def\colorgray#1{\color[gray]{#1}}%
      \expandafter\def\csname LTw\endcsname{\color{white}}%
      \expandafter\def\csname LTb\endcsname{\color{black}}%
      \expandafter\def\csname LTa\endcsname{\color{black}}%
      \expandafter\def\csname LT0\endcsname{\color{black}}%
      \expandafter\def\csname LT1\endcsname{\color{black}}%
      \expandafter\def\csname LT2\endcsname{\color{black}}%
      \expandafter\def\csname LT3\endcsname{\color{black}}%
      \expandafter\def\csname LT4\endcsname{\color{black}}%
      \expandafter\def\csname LT5\endcsname{\color{black}}%
      \expandafter\def\csname LT6\endcsname{\color{black}}%
      \expandafter\def\csname LT7\endcsname{\color{black}}%
      \expandafter\def\csname LT8\endcsname{\color{black}}%
    \fi
  \fi
    \setlength{\unitlength}{0.0500bp}%
    \ifx\gptboxheight\undefined%
      \newlength{\gptboxheight}%
      \newlength{\gptboxwidth}%
      \newsavebox{\gptboxtext}%
    \fi%
    \setlength{\fboxrule}{0.5pt}%
    \setlength{\fboxsep}{1pt}%
\begin{picture}(4896.00,2014.00)%
    \gplgaddtomacro\gplbacktext{%
      \csname LTb\endcsname%
      \put(417,302){\makebox(0,0)[r]{\strut{} 0k}}%
      \csname LTb\endcsname%
      \put(417,554){\makebox(0,0)[r]{\strut{}200k}}%
      \csname LTb\endcsname%
      \put(417,805){\makebox(0,0)[r]{\strut{}400k}}%
      \csname LTb\endcsname%
      \put(417,1057){\makebox(0,0)[r]{\strut{}600k}}%
      \csname LTb\endcsname%
      \put(417,1308){\makebox(0,0)[r]{\strut{}800k}}%
      \csname LTb\endcsname%
      \put(417,1560){\makebox(0,0)[r]{\strut{}1000k}}%
      \csname LTb\endcsname%
      \put(417,1811){\makebox(0,0)[r]{\strut{}1200k}}%
      \csname LTb\endcsname%
      \put(739,182){\makebox(0,0){\strut{}$2$}}%
      \csname LTb\endcsname%
      \put(1239,182){\makebox(0,0){\strut{}$4$}}%
      \csname LTb\endcsname%
      \put(1739,182){\makebox(0,0){\strut{}$6$}}%
      \csname LTb\endcsname%
      \put(2239,182){\makebox(0,0){\strut{}$8$}}%
    }%
    \gplgaddtomacro\gplfronttext{%
      \csname LTb\endcsname%
      \put(-75,1056){\rotatebox{-270}{\makebox(0,0){\strut{}Operations / sec}}}%
      \put(1364,2){\makebox(0,0){\strut{}\# cores}}%
      \put(1364,1991){\makebox(0,0){\strut{}(a) Throughput - 90\% reads}}%
      \csname LTb\endcsname%
      \put(1744,730){\makebox(0,0)[r]{\strut{}DRAM}}%
      \csname LTb\endcsname%
      \put(1744,580){\makebox(0,0)[r]{\strut{}Redo-opt}}%
      \csname LTb\endcsname%
      \put(1744,430){\makebox(0,0)[r]{\strut{}Redo}}%
    }%
    \gplgaddtomacro\gplbacktext{%
      \csname LTb\endcsname%
      \put(2706,302){\makebox(0,0)[r]{\strut{} 0k}}%
      \csname LTb\endcsname%
      \put(2706,554){\makebox(0,0)[r]{\strut{}200k}}%
      \csname LTb\endcsname%
      \put(2706,805){\makebox(0,0)[r]{\strut{}400k}}%
      \csname LTb\endcsname%
      \put(2706,1057){\makebox(0,0)[r]{\strut{}600k}}%
      \csname LTb\endcsname%
      \put(2706,1308){\makebox(0,0)[r]{\strut{}800k}}%
      \csname LTb\endcsname%
      \put(2706,1560){\makebox(0,0)[r]{\strut{}1000k}}%
      \csname LTb\endcsname%
      \put(2706,1811){\makebox(0,0)[r]{\strut{}1200k}}%
      \csname LTb\endcsname%
      \put(3028,182){\makebox(0,0){\strut{}$2$}}%
      \csname LTb\endcsname%
      \put(3528,182){\makebox(0,0){\strut{}$4$}}%
      \csname LTb\endcsname%
      \put(4027,182){\makebox(0,0){\strut{}$6$}}%
      \csname LTb\endcsname%
      \put(4527,182){\makebox(0,0){\strut{}$8$}}%
    }%
    \gplgaddtomacro\gplfronttext{%
      \csname LTb\endcsname%
      \put(3652,2){\makebox(0,0){\strut{}\# cores}}%
      \put(3652,1991){\makebox(0,0){\strut{}(b) Throughput - 50\% reads}}%
      \csname LTb\endcsname%
      \put(4032,730){\makebox(0,0)[r]{\strut{}Undo-opt}}%
      \csname LTb\endcsname%
      \put(4032,580){\makebox(0,0)[r]{\strut{}Undo}}%
      \csname LTb\endcsname%
      \put(4032,430){\makebox(0,0)[r]{\strut{}COW}}%
    }%
    \gplbacktext
    \put(-1000,-1000){\includegraphics{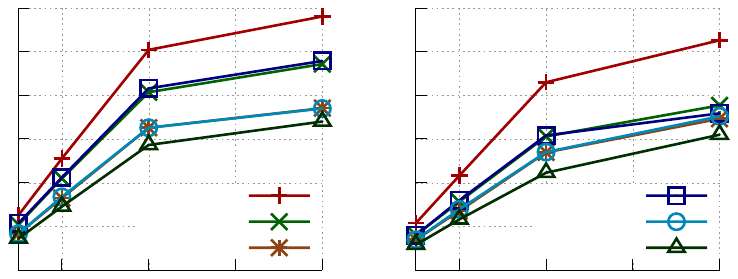}}%
    \gplfronttext
  \end{picture}%
\endgroup

%% file: data/app-sqlite.tex
\begingroup
\scriptsize
  \makeatletter
  \providecommand\color[2][]{%
    \GenericError{(gnuplot) \space\space\space\@spaces}{%
      Package color not loaded in conjunction with
      terminal option `colourtext'%
    }{See the gnuplot documentation for explanation.%
    }{Either use 'blacktext' in gnuplot or load the package
      color.sty in LaTeX.}%
    \renewcommand\color[2][]{}%
  }%
  \providecommand\includegraphics[2][]{%
    \GenericError{(gnuplot) \space\space\space\@spaces}{%
      Package graphicx or graphics not loaded%
    }{See the gnuplot documentation for explanation.%
    }{The gnuplot epslatex terminal needs graphicx.sty or graphics.sty.}%
    \renewcommand\includegraphics[2][]{}%
  }%
  \providecommand\rotatebox[2]{#2}%
  \@ifundefined{ifGPcolor}{%
    \newif\ifGPcolor
    \GPcolortrue
  }{}%
  \@ifundefined{ifGPblacktext}{%
    \newif\ifGPblacktext
    \GPblacktextfalse
  }{}%
  \let\gplgaddtomacro\g@addto@macro
  \gdef\gplbacktext{}%
  \gdef\gplfronttext{}%
  \makeatother
  \ifGPblacktext
    \def\colorrgb#1{}%
    \def\colorgray#1{}%
  \else
    \ifGPcolor
      \def\colorrgb#1{\color[rgb]{#1}}%
      \def\colorgray#1{\color[gray]{#1}}%
      \expandafter\def\csname LTw\endcsname{\color{white}}%
      \expandafter\def\csname LTb\endcsname{\color{black}}%
      \expandafter\def\csname LTa\endcsname{\color{black}}%
      \expandafter\def\csname LT0\endcsname{\color[rgb]{1,0,0}}%
      \expandafter\def\csname LT1\endcsname{\color[rgb]{0,1,0}}%
      \expandafter\def\csname LT2\endcsname{\color[rgb]{0,0,1}}%
      \expandafter\def\csname LT3\endcsname{\color[rgb]{1,0,1}}%
      \expandafter\def\csname LT4\endcsname{\color[rgb]{0,1,1}}%
      \expandafter\def\csname LT5\endcsname{\color[rgb]{1,1,0}}%
      \expandafter\def\csname LT6\endcsname{\color[rgb]{0,0,0}}%
      \expandafter\def\csname LT7\endcsname{\color[rgb]{1,0.3,0}}%
      \expandafter\def\csname LT8\endcsname{\color[rgb]{0.5,0.5,0.5}}%
    \else
      \def\colorrgb#1{\color{black}}%
      \def\colorgray#1{\color[gray]{#1}}%
      \expandafter\def\csname LTw\endcsname{\color{white}}%
      \expandafter\def\csname LTb\endcsname{\color{black}}%
      \expandafter\def\csname LTa\endcsname{\color{black}}%
      \expandafter\def\csname LT0\endcsname{\color{black}}%
      \expandafter\def\csname LT1\endcsname{\color{black}}%
      \expandafter\def\csname LT2\endcsname{\color{black}}%
      \expandafter\def\csname LT3\endcsname{\color{black}}%
      \expandafter\def\csname LT4\endcsname{\color{black}}%
      \expandafter\def\csname LT5\endcsname{\color{black}}%
      \expandafter\def\csname LT6\endcsname{\color{black}}%
      \expandafter\def\csname LT7\endcsname{\color{black}}%
      \expandafter\def\csname LT8\endcsname{\color{black}}%
    \fi
  \fi
    \setlength{\unitlength}{0.0500bp}%
    \ifx\gptboxheight\undefined%
      \newlength{\gptboxheight}%
      \newlength{\gptboxwidth}%
      \newsavebox{\gptboxtext}%
    \fi%
    \setlength{\fboxrule}{0.5pt}%
    \setlength{\fboxsep}{1pt}%
\begin{picture}(4608.00,2014.00)%
    \gplgaddtomacro\gplbacktext{%
      \csname LTb\endcsname%
      \put(516,240){\makebox(0,0)[r]{\strut{} 0}}%
      \csname LTb\endcsname%
      \put(516,512){\makebox(0,0)[r]{\strut{}400}}%
      \csname LTb\endcsname%
      \put(516,783){\makebox(0,0)[r]{\strut{}800}}%
      \csname LTb\endcsname%
      \put(516,1055){\makebox(0,0)[r]{\strut{}1200}}%
      \csname LTb\endcsname%
      \put(516,1326){\makebox(0,0)[r]{\strut{}1600}}%
      \csname LTb\endcsname%
      \put(516,1598){\makebox(0,0)[r]{\strut{}2000}}%
      \csname LTb\endcsname%
      \put(516,1869){\makebox(0,0)[r]{\strut{}2400}}%
      \csname LTb\endcsname%
      \put(873,120){\makebox(0,0){\strut{}Stock}}%
      \csname LTb\endcsname%
      \put(2299,120){\makebox(0,0){\strut{}PDOM-0}}%
      \csname LTb\endcsname%
      \put(3250,120){\makebox(0,0){\strut{}PDOM-1}}%
      \csname LTb\endcsname%
      \put(4201,120){\makebox(0,0){\strut{}PDOM-2}}%
    }%
    \gplgaddtomacro\gplfronttext{%
      \csname LTb\endcsname%
      \put(96,1054){\rotatebox{-270}{\makebox(0,0){\strut{}\# transactions / sec}}}%
      \csname LTb\endcsname%
      \put(1420,1793){\makebox(0,0)[r]{\strut{}In-memory}}%
      \csname LTb\endcsname%
      \put(1420,1643){\makebox(0,0)[r]{\strut{}Rollback}}%
      \csname LTb\endcsname%
      \put(2563,1793){\makebox(0,0)[r]{\strut{}WAL}}%
      \csname LTb\endcsname%
      \put(2563,1643){\makebox(0,0)[r]{\strut{}Undo}}%
      \csname LTb\endcsname%
      \put(3706,1793){\makebox(0,0)[r]{\strut{}Redo}}%
    }%
    \gplbacktext
    \put(-1000,-1000){\includegraphics{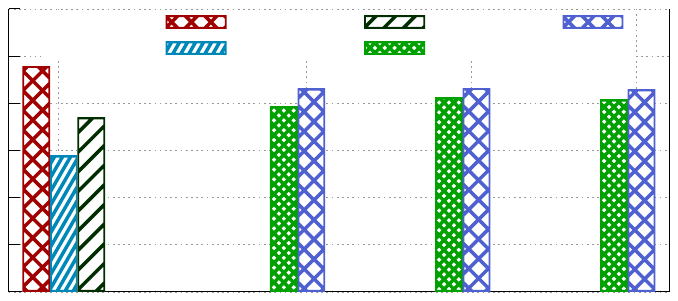}}%
    \gplfronttext
  \end{picture}%
\endgroup

%% file: data/app-memcached.tex
\begingroup
\scriptsize
  \makeatletter
  \providecommand\color[2][]{%
    \GenericError{(gnuplot) \space\space\space\@spaces}{%
      Package color not loaded in conjunction with
      terminal option `colourtext'%
    }{See the gnuplot documentation for explanation.%
    }{Either use 'blacktext' in gnuplot or load the package
      color.sty in LaTeX.}%
    \renewcommand\color[2][]{}%
  }%
  \providecommand\includegraphics[2][]{%
    \GenericError{(gnuplot) \space\space\space\@spaces}{%
      Package graphicx or graphics not loaded%
    }{See the gnuplot documentation for explanation.%
    }{The gnuplot epslatex terminal needs graphicx.sty or graphics.sty.}%
    \renewcommand\includegraphics[2][]{}%
  }%
  \providecommand\rotatebox[2]{#2}%
  \@ifundefined{ifGPcolor}{%
    \newif\ifGPcolor
    \GPcolortrue
  }{}%
  \@ifundefined{ifGPblacktext}{%
    \newif\ifGPblacktext
    \GPblacktextfalse
  }{}%
  \let\gplgaddtomacro\g@addto@macro
  \gdef\gplbacktext{}%
  \gdef\gplfronttext{}%
  \makeatother
  \ifGPblacktext
    \def\colorrgb#1{}%
    \def\colorgray#1{}%
  \else
    \ifGPcolor
      \def\colorrgb#1{\color[rgb]{#1}}%
      \def\colorgray#1{\color[gray]{#1}}%
      \expandafter\def\csname LTw\endcsname{\color{white}}%
      \expandafter\def\csname LTb\endcsname{\color{black}}%
      \expandafter\def\csname LTa\endcsname{\color{black}}%
      \expandafter\def\csname LT0\endcsname{\color[rgb]{1,0,0}}%
      \expandafter\def\csname LT1\endcsname{\color[rgb]{0,1,0}}%
      \expandafter\def\csname LT2\endcsname{\color[rgb]{0,0,1}}%
      \expandafter\def\csname LT3\endcsname{\color[rgb]{1,0,1}}%
      \expandafter\def\csname LT4\endcsname{\color[rgb]{0,1,1}}%
      \expandafter\def\csname LT5\endcsname{\color[rgb]{1,1,0}}%
      \expandafter\def\csname LT6\endcsname{\color[rgb]{0,0,0}}%
      \expandafter\def\csname LT7\endcsname{\color[rgb]{1,0.3,0}}%
      \expandafter\def\csname LT8\endcsname{\color[rgb]{0.5,0.5,0.5}}%
    \else
      \def\colorrgb#1{\color{black}}%
      \def\colorgray#1{\color[gray]{#1}}%
      \expandafter\def\csname LTw\endcsname{\color{white}}%
      \expandafter\def\csname LTb\endcsname{\color{black}}%
      \expandafter\def\csname LTa\endcsname{\color{black}}%
      \expandafter\def\csname LT0\endcsname{\color{black}}%
      \expandafter\def\csname LT1\endcsname{\color{black}}%
      \expandafter\def\csname LT2\endcsname{\color{black}}%
      \expandafter\def\csname LT3\endcsname{\color{black}}%
      \expandafter\def\csname LT4\endcsname{\color{black}}%
      \expandafter\def\csname LT5\endcsname{\color{black}}%
      \expandafter\def\csname LT6\endcsname{\color{black}}%
      \expandafter\def\csname LT7\endcsname{\color{black}}%
      \expandafter\def\csname LT8\endcsname{\color{black}}%
    \fi
  \fi
    \setlength{\unitlength}{0.0500bp}%
    \ifx\gptboxheight\undefined%
      \newlength{\gptboxheight}%
      \newlength{\gptboxwidth}%
      \newsavebox{\gptboxtext}%
    \fi%
    \setlength{\fboxrule}{0.5pt}%
    \setlength{\fboxsep}{1pt}%
\begin{picture}(10512.00,2014.00)%
    \gplgaddtomacro\gplbacktext{%
      \csname LTb\endcsname%
      \put(453,362){\makebox(0,0)[r]{\strut{} 0k}}%
      \csname LTb\endcsname%
      \put(453,692){\makebox(0,0)[r]{\strut{}50k}}%
      \csname LTb\endcsname%
      \put(453,1022){\makebox(0,0)[r]{\strut{}100k}}%
      \csname LTb\endcsname%
      \put(453,1353){\makebox(0,0)[r]{\strut{}150k}}%
      \csname LTb\endcsname%
      \put(453,1683){\makebox(0,0)[r]{\strut{}200k}}%
      \csname LTb\endcsname%
      \put(453,2013){\makebox(0,0)[r]{\strut{}250k}}%
      \csname LTb\endcsname%
      \put(767,242){\makebox(0,0){\strut{}$2$}}%
      \csname LTb\endcsname%
      \put(1251,242){\makebox(0,0){\strut{}$4$}}%
      \csname LTb\endcsname%
      \put(1736,242){\makebox(0,0){\strut{}$6$}}%
      \csname LTb\endcsname%
      \put(2220,242){\makebox(0,0){\strut{}$8$}}%
    }%
    \gplgaddtomacro\gplfronttext{%
      \csname LTb\endcsname%
      \put(33,1187){\rotatebox{-270}{\makebox(0,0){\strut{}Operations / sec}}}%
      \put(1372,62){\makebox(0,0){\strut{}\# cores}}%
      \put(1372,2193){\makebox(0,0){\strut{}(a) Throughput - 90\% reads}}%
      \csname LTb\endcsname%
      \put(1725,815){\makebox(0,0)[r]{\strut{}Undo/PDOM-0}}%
      \csname LTb\endcsname%
      \put(1725,665){\makebox(0,0)[r]{\strut{}Redo/PDOM-0}}%
      \csname LTb\endcsname%
      \put(1725,515){\makebox(0,0)[r]{\strut{}Undo/PDOM-1}}%
    }%
    \gplgaddtomacro\gplbacktext{%
      \csname LTb\endcsname%
      \put(2779,362){\makebox(0,0)[r]{\strut{} 0k}}%
      \csname LTb\endcsname%
      \put(2779,750){\makebox(0,0)[r]{\strut{}20k}}%
      \csname LTb\endcsname%
      \put(2779,1139){\makebox(0,0)[r]{\strut{}40k}}%
      \csname LTb\endcsname%
      \put(2779,1527){\makebox(0,0)[r]{\strut{}60k}}%
      \csname LTb\endcsname%
      \put(2779,1916){\makebox(0,0)[r]{\strut{}80k}}%
      \csname LTb\endcsname%
      \put(3093,242){\makebox(0,0){\strut{}$2$}}%
      \csname LTb\endcsname%
      \put(3577,242){\makebox(0,0){\strut{}$4$}}%
      \csname LTb\endcsname%
      \put(4062,242){\makebox(0,0){\strut{}$6$}}%
      \csname LTb\endcsname%
      \put(4546,242){\makebox(0,0){\strut{}$8$}}%
    }%
    \gplgaddtomacro\gplfronttext{%
      \csname LTb\endcsname%
      \put(3698,62){\makebox(0,0){\strut{}\# cores}}%
      \put(3698,2193){\makebox(0,0){\strut{}(b) Throughput - 50\% reads}}%
      \csname LTb\endcsname%
      \put(4051,773){\makebox(0,0)[r]{\strut{}Redo/PDOM-1}}%
      \csname LTb\endcsname%
      \put(4051,623){\makebox(0,0)[r]{\strut{}Undo/PDOM-2}}%
      \csname LTb\endcsname%
      \put(4051,473){\makebox(0,0)[r]{\strut{}Redo/PDOM-2}}%
    }%
    \gplgaddtomacro\gplbacktext{%
      \csname LTb\endcsname%
      \put(5105,362){\makebox(0,0)[r]{\strut{} 0}}%
      \csname LTb\endcsname%
      \put(5105,692){\makebox(0,0)[r]{\strut{}50}}%
      \csname LTb\endcsname%
      \put(5105,1022){\makebox(0,0)[r]{\strut{}100}}%
      \csname LTb\endcsname%
      \put(5105,1353){\makebox(0,0)[r]{\strut{}150}}%
      \csname LTb\endcsname%
      \put(5105,1683){\makebox(0,0)[r]{\strut{}200}}%
      \csname LTb\endcsname%
      \put(5105,2013){\makebox(0,0)[r]{\strut{}250}}%
      \csname LTb\endcsname%
      \put(5078,-70){\rotatebox{45}{\makebox(0,0)[l]{\strut{}Undo/P0}}}%
      \csname LTb\endcsname%
      \put(5370,-70){\rotatebox{45}{\makebox(0,0)[l]{\strut{}Redo/P0}}}%
      \csname LTb\endcsname%
      \put(5662,-70){\rotatebox{45}{\makebox(0,0)[l]{\strut{}Undo/P1}}}%
      \csname LTb\endcsname%
      \put(5954,-70){\rotatebox{45}{\makebox(0,0)[l]{\strut{}Redo/P1}}}%
      \csname LTb\endcsname%
      \put(6246,-70){\rotatebox{45}{\makebox(0,0)[l]{\strut{}Undo/P2}}}%
      \csname LTb\endcsname%
      \put(6538,-70){\rotatebox{45}{\makebox(0,0)[l]{\strut{}Redo/P2}}}%
    }%
    \gplgaddtomacro\gplfronttext{%
      \csname LTb\endcsname%
      \put(4757,1187){\rotatebox{-270}{\makebox(0,0){\strut{}Transactions latency (usec)}}}%
      \put(6024,2193){\makebox(0,0){\strut{}(c) Get / Put latency - 90\% reads}}%
      \csname LTb\endcsname%
      \put(6376,1938){\makebox(0,0)[r]{\strut{}Get}}%
      \csname LTb\endcsname%
      \put(6376,1788){\makebox(0,0)[r]{\strut{}Put}}%
    }%
    \gplgaddtomacro\gplbacktext{%
      \csname LTb\endcsname%
      \put(7430,362){\makebox(0,0)[r]{\strut{} 0}}%
      \csname LTb\endcsname%
      \put(7430,692){\makebox(0,0)[r]{\strut{}100}}%
      \csname LTb\endcsname%
      \put(7430,1022){\makebox(0,0)[r]{\strut{}200}}%
      \csname LTb\endcsname%
      \put(7430,1353){\makebox(0,0)[r]{\strut{}300}}%
      \csname LTb\endcsname%
      \put(7430,1683){\makebox(0,0)[r]{\strut{}400}}%
      \csname LTb\endcsname%
      \put(7430,2013){\makebox(0,0)[r]{\strut{}500}}%
      \csname LTb\endcsname%
      \put(7403,-70){\rotatebox{45}{\makebox(0,0)[l]{\strut{}Undo/P0}}}%
      \csname LTb\endcsname%
      \put(7695,-70){\rotatebox{45}{\makebox(0,0)[l]{\strut{}Redo/P0}}}%
      \csname LTb\endcsname%
      \put(7987,-70){\rotatebox{45}{\makebox(0,0)[l]{\strut{}Undo/P1}}}%
      \csname LTb\endcsname%
      \put(8280,-70){\rotatebox{45}{\makebox(0,0)[l]{\strut{}Redo/P1}}}%
      \csname LTb\endcsname%
      \put(8572,-70){\rotatebox{45}{\makebox(0,0)[l]{\strut{}Undo/P2}}}%
      \csname LTb\endcsname%
      \put(8864,-70){\rotatebox{45}{\makebox(0,0)[l]{\strut{}Redo/P2}}}%
    }%
    \gplgaddtomacro\gplfronttext{%
      \csname LTb\endcsname%
      \put(8349,2193){\makebox(0,0){\strut{}(d) Get / Put latency - 50\% reads}}%
    }%
    \gplbacktext
    \put(-1000,-1000){\includegraphics{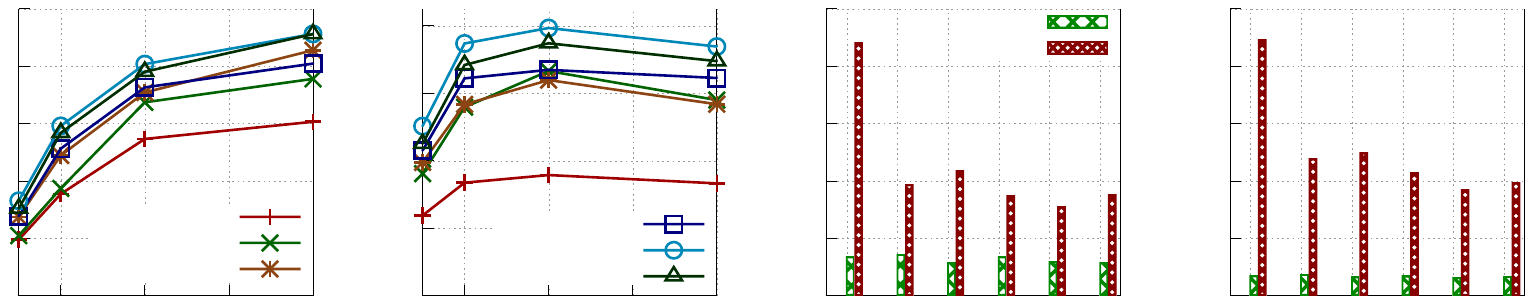}}%
    \gplfronttext
  \end{picture}%
\endgroup

%% file: related-work.tex
\section{Related Work}
\label{sec:related-work}

While most early work on persistent memory transaction runtimes
ignores persist barrier
overheads~\cite{bridge15,chakrabarti14,coburn11,pmem-io}, a growing
number of efforts~\cite{giles15,kolli16,lu16,volos11} is addressing
the problem in different ways.  

Volos et al.~\cite{volos11} implement redo logging transactions in
their Mnemosyne runtime.
Giles et al.~\cite{giles15} go further with a ``lazy'' cleanup
proposal that moves redo log application and cleanup to a background
thread.  They additionally add a DRAM-based aliasing mechanism to
cache persistent objects in the faster DRAM.  We experimented with
this scheme in our framework and found that the alias table lookup
induced hardware cache misses offset any gains provided by the faster
DRAM cache.
Lu et al.~\cite{lu16} optimize out Mnemosyne transactions' last
persist barrier, and propose a full processor cache flush as a
technique to checkpoint committed transactions' results.
Kolli et al.~\cite{kolli16} propose an undo logging based transaction
runtime that introduces just four persist barriers per transaction,
assuming that transactions know the data they need to modify in
advance.

In contrast to these works that focus on one specific transaction
runtime implementation, we perform a far more comprehensive analysis
of the design space. Our analysis not only considers workload
characteristics but also performance implications of persistence
domains. In addition, we also report implications of cache locality.

Moraru et al.~\cite{moraru13} present a memory allocator optimized for
wear leveling, which can plug easily into our allocation log
technique.  Volos et al.~\cite{volos11} presented an allocator that
uses the transaction's redo log to track persistent memory
allocations.  This is similar to our memory allocator.  However
we splice out the allocation records into a separate allocation log
that lets us use it in undo and COW transaction runtimes.

Work similar to ours has emerged in the in-memory database
setting~\cite{arulraj15}, where the authors compare database
transactions based on ``in-place'' updates (similar to our undo/redo
logging runtimes), copy-on-write, and write-ahead
logging~\cite{mohan92} implementations.  While their results align
with ours -- in-place updates tend to dominate over the other two
approaches -- the two settings are significantly different.  Their
runtimes are designed to optimize database processing, with hand
optimized implementations of core database data structures, whereas
ours are much lower level runtimes developed to track individual loads
and stores to arbitrary data structures hosted in persistent memory.

Different memory persistency models have been proposed in academia
over the past few years~\cite{condit09,joshi15,pelley14,ren15}.
However, there seems to be a convergence emerging on the thread-local
\emph{epoch persistency} model~\cite{condit09,pelley14} with Intel's
recent deprecation of the \pcommit\ instruction from future Intel
processors.  Our work applies to all these persistency models.

%% file: conclusion.tex
\section{Conclusion}
\label{sec:conclusion}

We presented a new taxonomy of persistence domains based on our
observations of support for persistent memory in past and future
systems.  We also presented three transaction runtime systems based on
undo logging, redo logging and copy-on-write implementations of
transactional writes.  Our runtimes are designed with the goal to
reduce persist barriers needed in a transaction.  Our allocator is
also designed with the same goal and plugs into all our transaction
runtimes.  Our comprehensive microbenchmark-based evaluation combs
through the read/write mix spectrum as well as persistence domain
choices showing that there exists a complex interplay between several
factors that contribute to performance of transaction runtimes --
workload read/write access patterns, persistence domains, cache
locality, and transaction runtime bookkeeping overheads.  Our ``real
world'' workload evaluations nicely conform to our microbenchmark
analysis, and provide insights into influence of additional
complexities including networking overheads (our K-V store), and
synchronization in multi-threaded applications (\memcached).  While
COW transactions appear to be a nonstarter, the choice between redo
and undo logging based runtimes is non-trivial, and needs to be
informed by the various parameters pertinent to the workload and the
enclosing system's support for persistence.